\documentclass[12pt]{article}

\usepackage{newtxtext,newtxmath}
\usepackage{graphicx}

\usepackage[letterpaper,margin=1in]{geometry}

\renewenvironment{abstract}
	{\quotation}
	{\endquotation}

\date{}

\makeatletter
\renewcommand{\fnum@figure}{\textbf{Figure \thefigure}}
\renewcommand{\fnum@table}{\textbf{Table \thetable}}
\makeatother

\usepackage{scicite}

\usepackage{url}

\newcommand{\ldl}{$\lambda/\Delta\lambda$}
\newcommand{\micron}{$\upmu$m}
\newcommand{\teff}{T$_\mathrm{eff}$}
\newcommand{\logg}{$\log_\mathrm{10}{\mathrm{g}}$}

\newcommand{\logl}{$\log_\mathrm{10}\left(L_\mathrm{bol}/L_\odot\right)$}
\newcommand{\kzz}{$\log_\mathrm{10}{\left(\kappa_\mathrm{zz}/\mathrm{cm^2~s^{-1}}\right)}$}

\newcommand{\vsini}{$v\sin{\mathrm{i}}$}
\newcommand{\hh}{H$_2$}
\newcommand{\hho}{H$_2$O}
\newcommand{\hhs}{H$_2$S}
\newcommand{\coo}{CO$_2$}
\newcommand{\co}{CO}
\newcommand{\chhhh}{CH$_4$}
\newcommand{\phhh}{PH$_3$}
\newcommand{\nhhh}{NH$_3$}
\newcommand{\msun}{M$_\odot$}

\newcommand{\mjup}{M$_\mathrm{Jup}$}
\newcommand{\rjup}{R$_\mathrm{Jup}$}
\newcommand{\name}{Wolf~1130C}
\newcommand{\adp}{NH$_4$H$_2$PO$_4$}
\newcommand\degree{\mbox{$^{\circ}$}}% 
\newcommand\farcs{\mbox{$.\!\!^{\prime\prime}$}}% 

\def\scititle{
	Observation of undepleted phosphine in the atmosphere of a low-temperature brown dwarf
}
\title{\bfseries \boldmath \scititle}

\author{
Adam J.\ Burgasser$^{1\ast}$,
Eileen C.\ Gonzales$^{2}$,
Samuel A.\ Beiler$^{3}$,
Channon Visscher$^{4,5}$,
\and Ben Burningham$^{6}$,
Gregory N.\ Mace$^{7}$,
Jacqueline K.\ Faherty$^{8}$,
Zenghua Zhang$^{9,10}$,
\and Clara Sousa-Silva$^{11,12}$,
Nicolas Lodieu$^{13,14}$,
Stanimir A.\ Metchev$^{15}$,
\and Aaron Meisner$^{16}$,
Michael Cushing$^{3}$,
Adam C.\ Schneider$^{17}$,
Genaro Suarez$^{8}$,
\and Chih-Chun Hsu$^{18}$,
Roman Gerasimov$^{19}$,
Christian Aganze$^{20}$,
\and Christopher A. Theissen$^{1}$
\and\small$^{1}$Department of Astronomy \& Astrophysics, University of California, San Diego, La Jolla, CA, USA
\and\small$^{2}$Department of Physics and Astronomy, San Francisco State University, San Francisco, CA, USA
\and\small$^{3}$Ritter Astrophysical Research Center, Department of Physics \& Astronomy, University of Toledo,
\and\small{~~~~~~~~~~~~Toledo, OH, USA~~~~~~~~~~}
\and\small$^{4}$Chemistry \& Planetary Sciences, Dordt University, Sioux Center, IA, USA
\and\small$^{5}$Center for Exoplanetary Systems, Space Science Institute, Boulder, CO, USA
\and\small$^{6}$Centre for Astrophysics Research, Department of Physics, Astronomy and Mathematics,
\and\small{University of Hertfordshire, Hatfield, UK}
\and\small$^{7}$Department of Astronomy, University of Texas at Austin, Austin, TX, USA
\and\small$^{8}$Department of Astrophysics, American Museum of Natural History, New York, NY, USA
\and\small$^{9}$School of Astronomy and Space Science, Nanjing University, Nanjing, China
\and\small$^{10}$Key Laboratory of Modern Astronomy and Astrophysics, Nanjing University, Ministry of Education,
\and\small{~~~~~~~~~~~~~~~~~~~~~~~~~~~~~~~Nanjing, China~~~~~~~~~~~~~~~~~~~~~~~~~~~~~~~}
\and\small$^{11}$Bard College, Annandale-on-Hudson, NY, USA
\and\small$^{12}$Institute of Astrophysics and Space Sciences, Porto, Portugal
\and\small$^{13}$Instituto de Astrof\'isica de Canarias, La Laguna, Tenerife, Spain
\and\small$^{14}$Departamento de Astrof\'isica, Universidad de La Laguna, La Laguna, Tenerife, Spain
\and\small$^{15}$Department of Physics \& Astronomy, Western University and Institute for Earth and Space Exploration,
\and\small{~~~~~~~~~~London, ON, Canada~~~~~~~~~~}
\and\small$^{16}$National Science Foundation's National Optical-Infrared Astronomy Research Laboratory, Tucson, AZ, USA
\and\small$^{17}$United States Naval Observatory, Flagstaff Station, Flagstaff, AZ, USA
\and\small$^{18}$Center for Interdisciplinary Exploration and Research in Astrophysics, Northwestern University,
\and\small{~~~~~~~~~~Evanston, IL, USA~~~~~~~~~~}
\and\small$^{19}$Department of Physics and Astronomy, University of Notre Dame, Notre Dame, IN, USA
\and\small$^{20}$Kavli Institute for Particle Astrophysics \& Cosmology, Stanford University, Stanford, CA, USA
\and\small$^\ast$Corresponding author. Email: aburgasser@ucsd.edu
}

\begin{document} 

% Insert the title and author list
\maketitle

\begin{abstract} \bfseries \boldmath
The atmospheres of low-temperature brown dwarfs and gas giant planets are expected to contain the phosphine molecule, {\phhh}. 
However, previous observations have shown much lower abundances of this molecule than predicted by atmospheric chemistry models.
We report JWST spectroscopic observations of phosphine  in the atmosphere of the brown dwarf {\name}. 
%JWST/NIRSpec spectroscopy reveals the fundamental $\nu_1+\nu_2$ stretch modes of 
Multiple absorption lines due to phosphine are detected around 4.3~${\upmu}$m, from which we calculate a 
phosphine abundance of 0.100$\pm$0.009 parts per million. 
This abundance is consistent with disequilibrium atmospheric chemistry models that reproduce the phosphine abundances in Jupiter and Saturn, and is much higher than abundances previously reported for other brown dwarfs or exoplanets. 
%Our analysis demonstrates an approach for studying phosphorus enrichment throughout the Milky Way.
This difference may be related to the low abundance of heavy elements in {\name}.
%and possibly contamination from the evolved white dwarf component Wolf~1130B.
\end{abstract}

% The first paragraph of any Science paper does NOT have a heading
% Nor is it indented
\noindent
Phosphorus and phosphorous molecules have been observed 
in the atmospheres of Saturn 
\cite{1975ApJ...202L..55B}
and Jupiter 
\cite{1976ApJ...207.1002R},
in the interstellar medium 
\cite{1978ApJ...219..861J}, 
and in the circumstellar material around evolved stars 
% DROPPED \cite{1990A&A...230L...9G,
\cite{2014ApJ...790L..27A}.
The molecular hydride phosphine ({\phhh}) has been proposed as a potential biosignature for exoplanet atmospheres \cite{2020AsBio..20..235S}.
Phosphorus is primarily synthesized in the interiors of massive stars through the slow neutron capture process
\cite{1957RvMP...29..547B} 
and released into the interstellar medium when they explode as supernovae.
The abundance of phosphorus within the Milky Way therefore traces the history of massive stars over the lifetime of the Galaxy
\cite{2012A&A...540A..33C,2022AJ....164...61M}.

In the Sun, phosphorus has an abundance of about 0.3 parts per million (ppm) measured relative to hydrogen, the most abundant element \cite{2021A&A...653A.141A}.
Spectroscopic studies of bright giant 
stars at 
ultraviolet \cite{2014ApJ...797...69R}
and infrared wavelengths 
\cite{2011A&A...532A..98C,2022AJ....164...61M} 
have shown that phosphorus can be 2 to 3 times more abundant than predicted by Milky Way evolution models
\cite{2012A&A...540A..33C,2014ApJ...796L..24J}. 
A distinct class of phosphorus-rich (P-rich) stars have also been identified, with 
abundances ten times the solar value \cite{2020NatCo..11.3759M}. 
These P-rich stars are ancient giant stars with low atmospheric metallicity 
(low abundances of elements heavier than helium).
Their origin is unclear; they might indicate that our understanding of phosphorus production 
through neutron capture 
processes is incomplete \cite{2024A&A...690A.262B}
%\cite{2019AJ....158..219M} DROPPED}
or that there are
additional sites of phosphorus synthesis \cite{2024ApJ...967L...1B}.

Phosphine {has} been predicted to be the primary phosphorous molecule in the low-temperature atmospheres of brown dwarfs and {gas giant} exoplanets
\cite{1996ApJ...472L..37F,2006ApJ...648.1181V}.
The abundances of PH$_3$ in the atmospheres of Jupiter and Saturn are 2 to 6 ppm, implying that phosphorus 
in those planets is enriched by factors of 5 to 16 compared to the Sun \cite{1982Icar...49..416D,1998JGR...10323001I,2009Icar..202..543F}.
However, observational searches for
{\phhh} in brown dwarf
%\cite{2016ApJ...826L..17S,2018ApJ...858...97M,2020AJ....160...63M,2024ApJ...962..177B,2024ApJ...973...60B,2024ApJ...977L..49R}
\cite{2016ApJ...826L..17S,2018ApJ...858...97M,2020AJ....160...63M,2024ApJ...973...60B}
and exoplanet atmospheres 
\cite{2023Natur.614..659R,2024Natur.630..836W,2024AJ....168...77W}
have provided only upper limits 
that are 100 times lower
than the abundances predicted by atmosphere models \cite{2024ApJ...973...60B}. 
{\phhh} has been detected in the atmosphere of the low-temperature 
brown dwarf WISE~0855$-$0714 with an abundance of 1 part per billion (ppb), 200 times lower than 
predicted by models, even accounting for this object's low metallicity \cite{2024ApJ...977L..49R}.
Observational searches for {\phhh} in the infrared spectra of cool atmospheres are hampered by a strong absorption band of  {\coo} which overlaps with the 4.3~{\micron} band of {\phhh}.
In contrast to {\phhh}, {\coo} is up to 5,000 times {more abundant} than predicted from atmosphere models 
\cite{2024ApJ...973...60B}.
% DROPPED: 2010ApJ...722..682Y
These discrepancies indicate that our understanding of phosphorus chemistry and chemical dynamics in cool atmospheres is incomplete.

\section*{The Wolf 1130ABC system}\label{sec:w1130c}

The Wolf~1130ABC system (= HIP~98905 = LHS~482 = GJ~781AB)
is a triple star system 17~parsecs (pc) from the Sun, composed of a low mass star (Wolf~1130A = GJ~781A) and white dwarf (Wolf~1130B) in a short-period (0.5~day) binary orbit, 
with a cold ($\approx$ 600~K), widely-separated ($\sim$3,000~astronomical units) brown dwarf tertiary, {\name} (= GJ~781B = WISE J200520.38+542433.9; \cite{1921AN....213...31W,2013ApJ...777...36M,2018ApJ...854..145M}).
Wolf~1130A has an iron abundance five times lower than the Sun (logarithmic measure of [Fe/H] = $-$0.70$\pm$0.12)
and is enriched in elements associated with massive star nucleosynthesis
\cite{2006PASP..118..218W,2018ApJ...854..145M}.
These abundance patterns, and the high velocity of the system relative to the Sun, indicate 
that it is part of the Milky Way's thick disk, which consists of old stars 
\cite{1983MNRAS.202.1025G}.
Previous spectral observations of {\name} have shown that it also has a 
low metallicity atmosphere \cite{2013ApJ...777...36M,2025ApJ...982...79B}, and our observations confirm that the source has a low bolometric luminosity, {\logl} = $-$6.047$\pm$0.003, consistent with a low-temperature brown dwarf \cite{supp}.
%results that are validated and extended in this work.

\section*{Observation of Phosphine in {\name}}\label{sec:phosphine}

We observed {\name} using the Near-Infrared Spectrograph (NIRSpec \cite{2022A&A...661A..80J}) 
and the Mid-Infrared Imager (MIRI \cite{2023PASP..135d8003W})
on JWST on 2024 Aug 30 \cite{supp}.
%, as part of program JWST-GO-04668 (PI Burgasser \cite{2024jwst.prop.4668B}). The Prism mode was used
%PRISM and G395H modes of NIRSpec were used with a fixed slit 
With NIRSpec, we used the Prism mode with the combined S200A1 0$\farcs$2-wide slit to obtain low-resolution spectra (resolving power {\ldl} $\approx$ 100) spanning the wavelength range 0.6 to 5.2~{\micron}, and 
the G395H grating mode with the S200A1 and S200A2 0$\farcs$2-wide slits to obtain moderate-resolution spectra ({\ldl} $\approx$ 3,000) spanning 3.8 to 5.2~{\micron}.
With MIRI, we used the Low Resolution Spectroscopy (LRS) mode with the 0$\farcs$51-wide slit to obtain low-resolution spectra ({\ldl} $\approx$ 100) spanning 4 to 14~{\micron}.
For the NIRSpec/Prism observations, four exposures totaling 997~s were obtained, resulting in a median signal-to-noise (S/N) of 300 at 4.1~{\micron}.
For the NIRSpec/G395H observations, four exposures totaling 1099~s were obtained, resulting in a median S/N of 75 at 4.1~{\micron}.
For the MIRI/LRS observations, two exposures totaling 555~s were obtained, resulting in a median S/N of 90 at 9~{\micron}.
%at each of two positions
%Further details on the acquisition and reduction of these data are provided in the Supplementary Materials.

Fig.~\ref{fig:spectra}A shows the NIRSpec/Prism spectrum of {\name}, compared to that of the brown dwarf ULAS J102940.52+093514.6 (hereafter ULAS~J1029+0935), which has a similar
effective temperature ({\teff} $\approx$ 700~K \cite{2024ApJ...973..107B}) but kinematics consistent with the Milky 
Way's thin disk population \cite{2013MNRAS.433..457B}.
% Ross~458C, {the latter being} another widely-separated, 
% %low-mass (M = 5-25~M$_{Jup}$), and 
% low-temperature ({\teff}=650--725~K) brown dwarf companion to a young (150--800~Myr) and metal-rich ([Fe/H] = +0.30$\pm$0.08) binary {star system}
%\cite{2010MNRAS.405.1140G,2010ApJ...725.1405B,2011MNRAS.414.3590B,2012ApJ...756..172M,2014AJ....147..160M,2023MNRAS.521.5761G}. 
%\cite{2010MNRAS.405.1140G,2011MNRAS.414.3590B,2014AJ....147..160M,2023MNRAS.521.5761G}. 
The spectra of both sources have similar molecular absorption features, which are common to low-temperature brown dwarfs, but they are clearly distinct in the 1.5 to 2.5~{\micron} region where {\name} is fainter and the
3 to 5~{\micron} region where {\name} is brighter than ULAS~J1029+0935. 

The spectrum of {\name} has an additional absorption feature centered at 4.3~{\micron} (Fig.~\ref{fig:spectra}A-B). We identify this feature as the blended $\nu_1$ and $\nu_3$ fundamental vibrational bands of {\phhh}, where the subscript indicates the vibrational quantum number. Those bands are split into P-, Q-, and R-branch rotational lines, with the main feature at 4.3~{\micron} being the blended Q-branch lines. In the higher-resolution spectrum (Fig.~\ref{fig:spectra}D), we also identify individual P- and R-branch lines in the 4.1 to 4.2~{\micron} and 4.35 to 4.45~{\micron} regions respectively, interspersed with lines of {\chhhh} and {\hho}.
%with the characteristic PQR-branch symmetric top pattern of 
% the blended $\nu_1$ and $\nu_3$ fundamental bands for {\phhh} \cite{2013JMoSp.288...28S}.
In contrast, the spectrum of ULAS~J1029+0935 shows strong {\coo} absorption in this region (Fig.~\ref{fig:spectra}C), which is absent in the spectrum of {\name}.
% The higher-resolution G395H spectrum of {\name} {reveals} the {\phhh} feature in greater detail, {showing} both blended Q-branch lines at 4.3~{\micron} and isolated R- and P- branch lines over 4.1--4.2~{\micron} and 4.35--4.45~{\micron}, interspersed among {\chhhh} and {\hho} transitions. 
%{The presence of {\phhh} in the atmosphere of {\name} is unambiguous on visual inspection, and further validated in spectral modeling.}
%These data provide the first robust detection of {\phhh} in the atmosphere of a source outside the Solar System.

\section*{Atmospheric Chemistry Modeling}\label{sec:modeling}

\begin{figure}[h]
\centering
 A \\\includegraphics[width=0.75\textwidth]{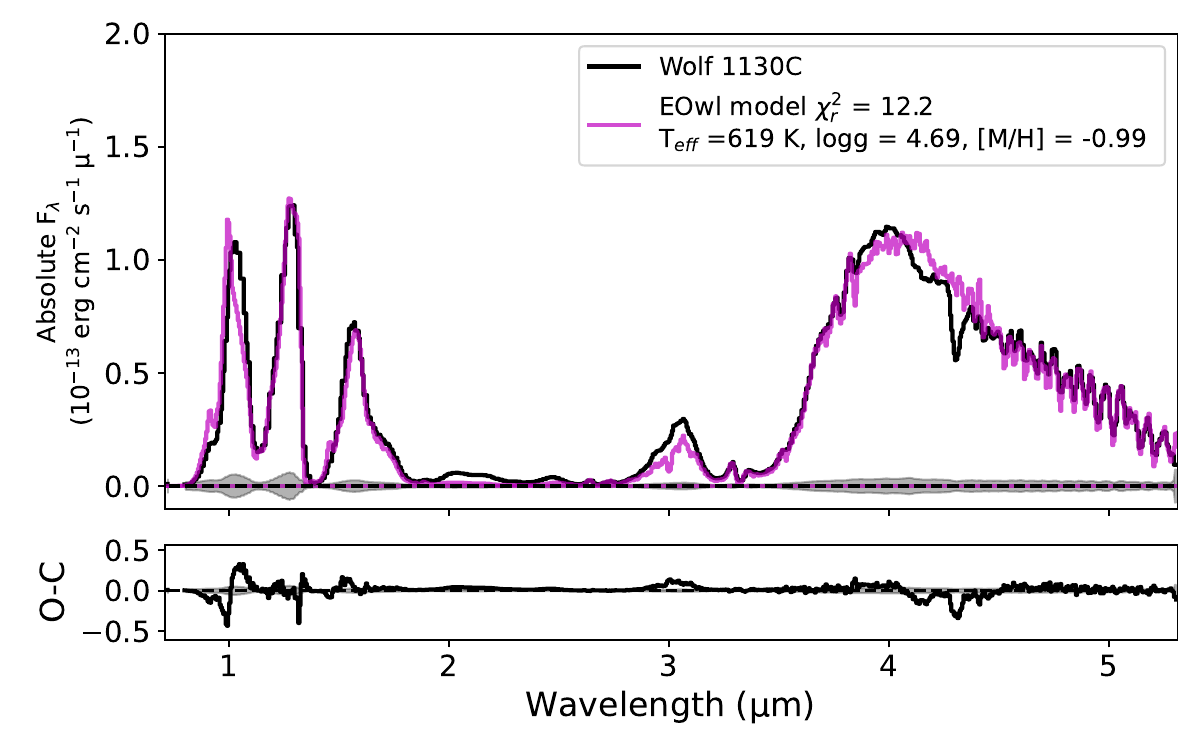} \\
 B \\\includegraphics[width=0.75\textwidth]{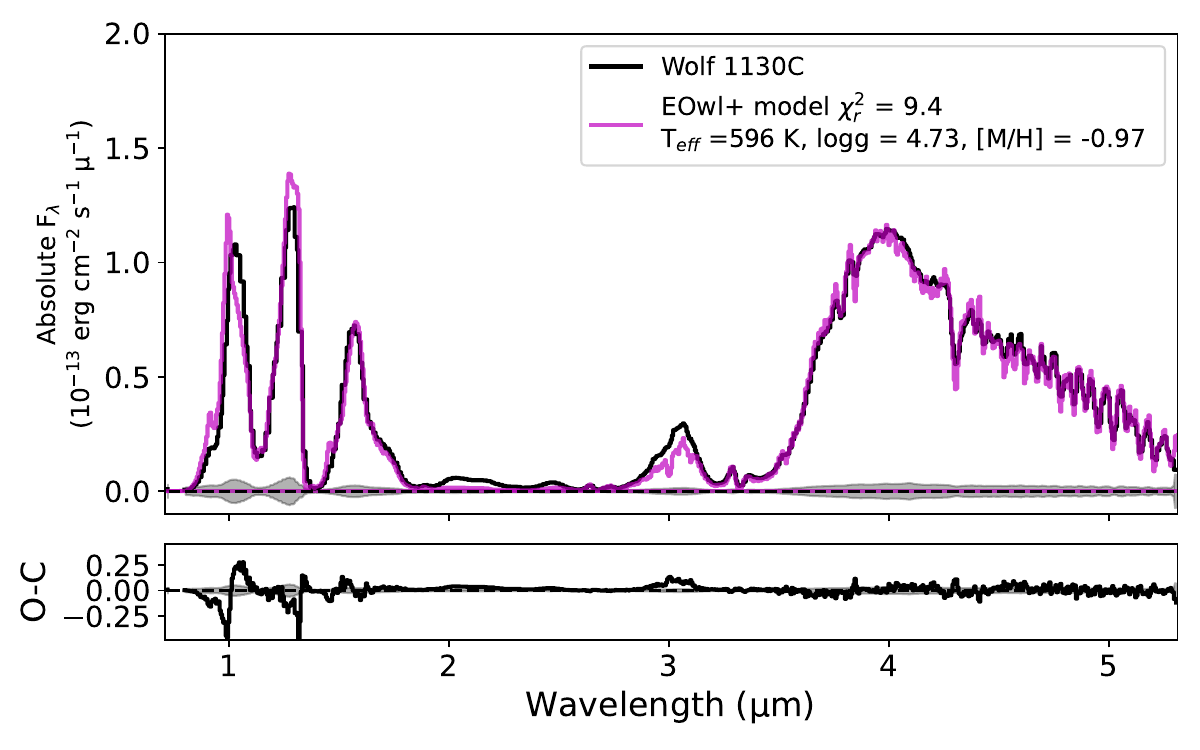} \\ 
\caption{{\bf
Sonora Elf-Owl models fitted to the spectrum of {\name}}.
(A) The observed spectrum fitted by
%Panel A displays fits using the published 
models with artificially suppressed {\phhh} abundances (EOwl \cite{2024ApJ...963...73M});
(B) the same but
%panel B displays fits 
using models with {\phhh} abundances
set by vertical mixing (EOwl+ \cite{2024ApJ...973...60B}). 
Both figures compare the absolute flux-calibrated spectrum (black lines) to the best-fitting model (magenta line) and {20} draws from the MCMC posterior chains (pale {magenta} lines) as an indication of the model uncertainty. Below each panel we show the difference (observed minus computed, or O-C) between the observed and best-fitting model spectra and the {$\pm$5$\sigma$} uncertainty of the former (grey bands).
The dashed black lines indicate zero flux.
}
\label{fig:gridfit}
\end{figure}

We compare the NIRSpec/Prism spectrum of {\name} 
to previously-published low-temperature atmospheric models  \cite{2024ApJ...963...73M,2024ApJ...973...60B} in Fig.~\ref{fig:gridfit} \cite{supp}.
%(see Supplementary Materials) demonstrates 
We find that the observed {\phhh} absorption bands
can be reproduced by models that include disequilibrium chemistry driven by vertical mixing of gas within the upper atmosphere.
This process is responsible for the {\phhh} enrichment observed in the atmospheres of Jupiter and Saturn \cite{1985ApJ...299.1067F,1994Icar..110..117F}.
Theoretical studies have predicted that vertical mixing affects the abundances of other molecules in low-temperature atmospheres, including
{\chhhh}, {\co}, {\coo}, and {\nhhh} \cite{1999ApJ...519L..85G,2006ApJ...647..552S}.
% DROPPED 2010ApJ...722..682Y
We examined modifications to the predicted 
atmospheric abundances of {\phhh} and {\coo} relative to disequilibrium chemistry models, accounting for the overall reduction of C and O elemental abundances of Wolf~1130A relative to the Sun  \cite{2024ApJ...973...60B}. 
We find that the line strengths of both absorption features 
%{($\leq2\times$)  
%and significant reduction in {\coo} abundances 
are consistent with model predictions \cite{supp},
%indicates a {\phhh} abundance consistent with that expected from vertical mixing from the deep atmosphere, with both 2$\times$ enhancement and decrement ruled out. 
%The absence of strong {\coo} absorption in the spectrum is also
%consistent with vertical mixing models, 
%when are accounted for .
in contrast to other brown dwarfs which show reduced abundances of {\phhh} and enhanced abundances of {\coo} relative to disequilibrium chemistry models \cite{2024ApJ...963...73M,2024ApJ...973...60B}. 
%The large {{\coo}} enhancements {inferred to be present} in the atmospheres of 
%{other} brown dwarfs \cite{2010ApJ...722..682Y,2024ApJ...973...60B} {is ruled out in the atmosphere of {\name}}.

\begin{figure}
\centering
%\includegraphics[width=0.6\textwidth]{Wolf1130C_NC_Pub_profile.pdf} \\
%A \\ 
\includegraphics[width=0.75\textwidth]{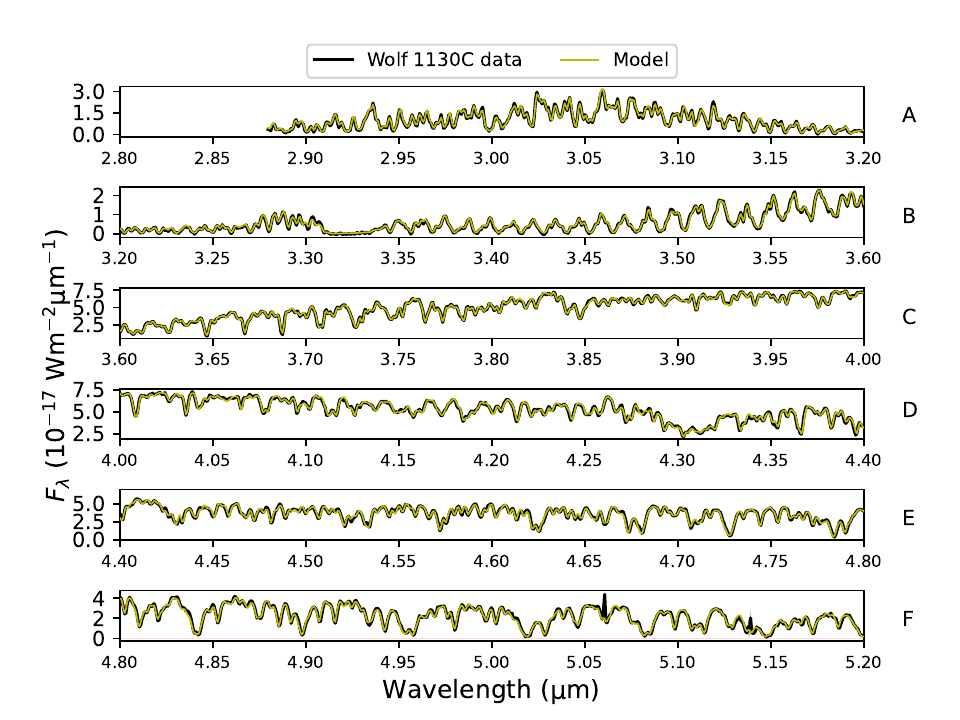}  \\ 
{\footnotesize G} \\ \includegraphics[width=0.55\textwidth]{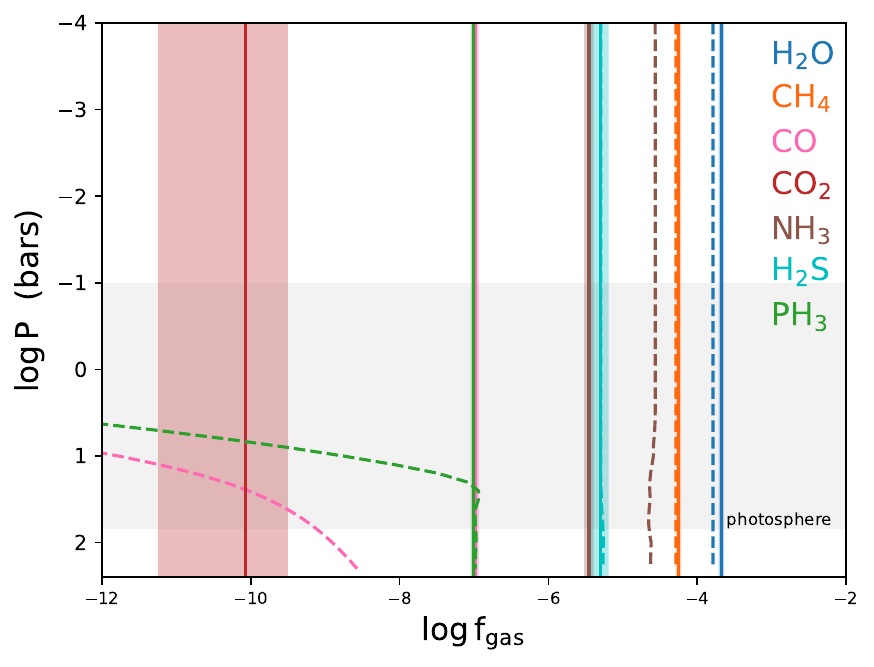} \\ 
\caption{{\bf {Retrieval analysis of the NIRSpec/G395H spectrum of {\name}}.}
%, and retrieved abundances compared to thermochemical equilibrium model predictions.}
{(A-F)} The G395H spectrum (black line) compared to the median likelihood
%and maximum likelihood (green line) 
atmosphere model from our retrieval analysis (mustard line). The two are nearly indistinguishable, with a reduced $\chi^2_r$ = 6.
Single pixel deviations at 5.06~{\micron} and 5.14~{\micron} are due to noise. 
{(G)} Retrieved abundances (solid vertical lines) for 
{\hho} {(blue)}, 
{\chhhh} {(orange)}, 
{\co} {(pink)}, 
{\coo} {(red)}, 
{\nhhh} {(brown)}, 
{\hhs} {(cyan)}, and 
{\phhh} {(green)} plotted as a function of logarithmic pressure (log~P) in the photosphere.
Shaded regions around each line indicate the $\pm$1$\sigma$ uncertainties on the retrieved abundances from the retrieval posterior distributions (Fig.~\ref{fig:retrieval3}).
Dashed lines are the predicted abundances for a [M/H] = $-$0.7, C/O = 0.26 thermochemical equilibrium model.
{\coo} is not visible in this plot as its equilibrium abundance is $<$10$^{-12}$ (Fig.~\ref{fig:chemistry2}) 
Grey shading indicates the approximate pressure region of the photosphere (visible layer of the atmosphere) in the 3 to 5~{\micron} region.
The retrieved abundances are assumed to be constant at all pressures.
% The thermochemical equilibrium model predicts {\phhh}, {\co}, and {\coo} (abundances 
% to be depleted in the upper photosphere.
}
\label{fig:retrieval1}
\end{figure}

We performed an abundance analysis of {\name} using the NIRSpec/G395H spectrum and the \textsc{Brewster} atmospheric retrieval framework
%We conducted a robust analysis of the atmosphere of {\name} using the moderate-resolution NIRSpec G395H spectrum and applying the {\it Brewster} retrieval framework 
%\cite{2017MNRAS.470.1177B,2021MNRAS.506.1944B,2021ApJ...923...19G}.
\cite{2017MNRAS.470.1177B,gonzales_2025_17082357,supp}.
%,Gonz20,2021ApJ...923...19G,2024Natur.628..511F}
%The Supplementary Materials provides {further details on} the fit procedure and inferred parameters. 
Fig.~\ref{fig:retrieval1}A shows the retrieved spectral model, which
%fit to the spectrum 
reproduces the
%2.9--5.2~{\micron} 
G395H spectrum. 
Fig.~\ref{fig:retrieval1}B compares the retrieved abundances of {\hho}, {\chhhh}, {\co}, {\coo}, {\nhhh}, {\hhs}, and {\phhh} to predictions of chemical equilibrium models for the metallicity and C/O ratio of the Wolf~1130 system \cite{2002Icar..155..393L,2006ApJ...648.1181V}.
The retrieval finds abundances of {\hho}, {\chhhh}, and {\hhs} that
%The abundances of {\phhh}, {\hhs}, and {\chhhh} 
are consistent with the predictions of equilibrium chemistry models, while abundances of {\co} and {\phhh} are consistent with vertical mixing of gas from deep below the photosphere.   
Our retrieved {\phhh} abundance of 0.100$\pm$0.009~ppm is 100 times larger than the {\phhh} abundance measured for WISE~J0855$-$0714 \cite{2024ApJ...977L..49R}. 
% Conversely, {\co}, like {\phhh}, has an abundance consistent with vertical mixing of gas from below the photosphere.
%assuming vertical mixing from the deep photosphere 
%at least at the base of the photosphere for {\phhh}.
%The abundance of {\nhhh} is 7 to 8 times below equilibrium chemistry model predictions, likely due to the vertical mixing of N$_2$ from the deep atmosphere \cite{2006ApJ...647..552S}.
Absorption features from {\coo} are only marginally detected, so its abundance is poorly constrained but likely less than 0.6~ppb. 
%{relative to chemical equilibrium and mixing models} is ambiguous {due to its marginal detection}.

From the retrieved molecular abundances, we infer that {\name} has a low metallicity, with a logarithmic metal abundance of [M/H] = $-$0.68$\pm$0.04 (20.9\%$\pm$1.9\%) relative to the Sun; 
%0.209$\pm$0.019 times the Sun's metal abundance; 
and a carbon/oxygen abundance ratio 0.26$\pm$0.01, also significantly less than the Sun. These values are
%Both of these values are 
consistent with the
composition of Wolf~1130A \cite{2018ApJ...854..145M} and other stars in the Milky Way's thick disk population \cite{2016A&A...594A..43H}.
% DROPPED \cite{2004A&A...414..931A}. 
Assuming these molecules contain the bulk of C, O, P, and S in the photosphere {of {\name},}
%, as expected from thermochemical equilibrium models, 
we calculate that it is enriched relative to the Sun 
%find {this source to be} modestly enriched
in phosphorus (1.7$\pm$0.2 times larger)
%[P/M] = +0.22$\pm$0.06), 
%sulfur ([S/M] = 1.6$\pm$0.6)
%+0.19$\pm$0.16),
and oxygen (2.1$\pm$0.3 times larger),
%[O/M] = +0.32$\pm$0.06),
but has a solar metal abundance fraction of carbon and sulfur.
%([C/M] = +0.04$\pm$0.06).
%The low {\nhhh} abundance ([N/M] = $-$0.60$\pm$0.13) suggests nitrogen depletion,
%although we cannot rule out a vertically-mixed disequilibrium N$_2$ reservoir.
These elemental abundance patterns match those of other metal-poor stars in the thick disk population \cite{2016A&A...594A..43H}. 
%{This analysis demonstrates the capability of retrieval modeling to measure} elemental abundances in {low-temperature} atmospheres.

\section*{Implications of Phosphine Detection}\label{sec:discussion}

\begin{figure}[h] % Do not use \begin{figure*}
%\centering
\hspace{0.5in}
\begin{minipage}{\textwidth}
\includegraphics[width=0.77\textwidth]{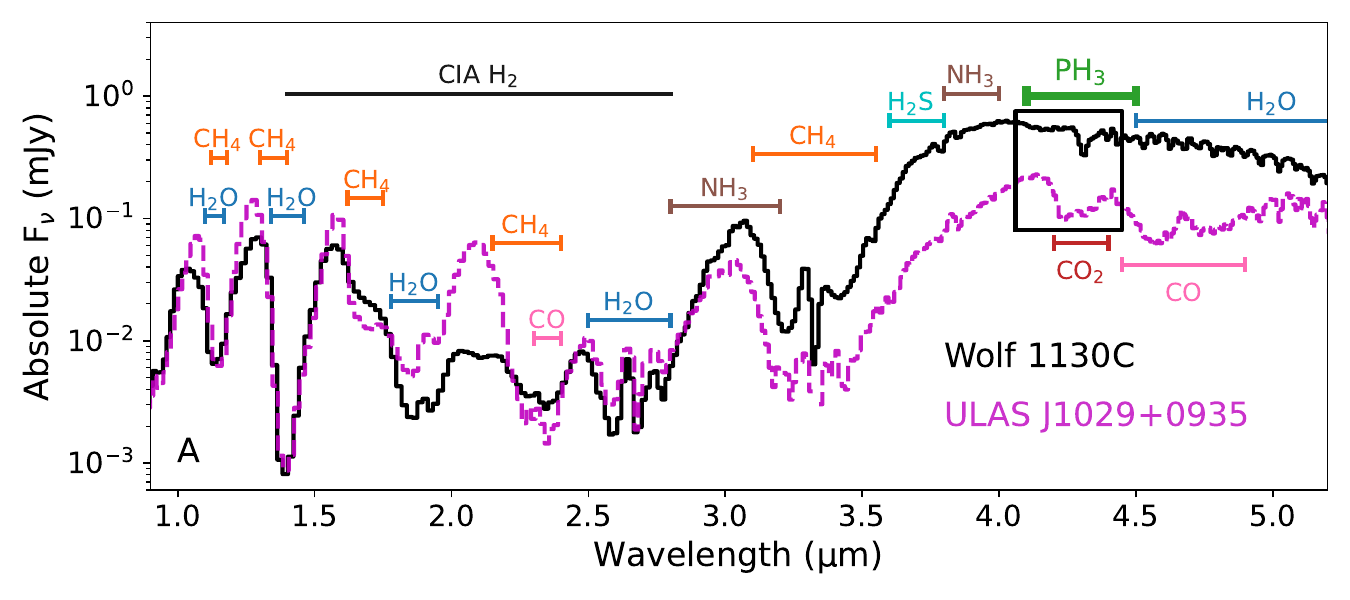} \\
\vspace{-0.2in}
~\includegraphics[width=0.375\textwidth]{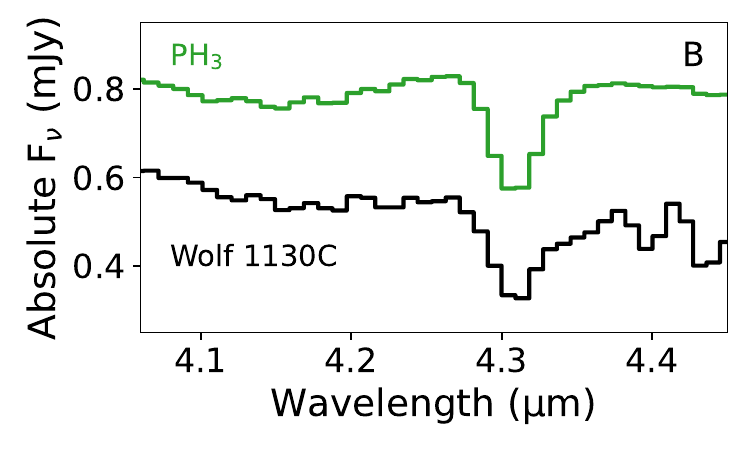} 
\includegraphics[width=0.375\textwidth]{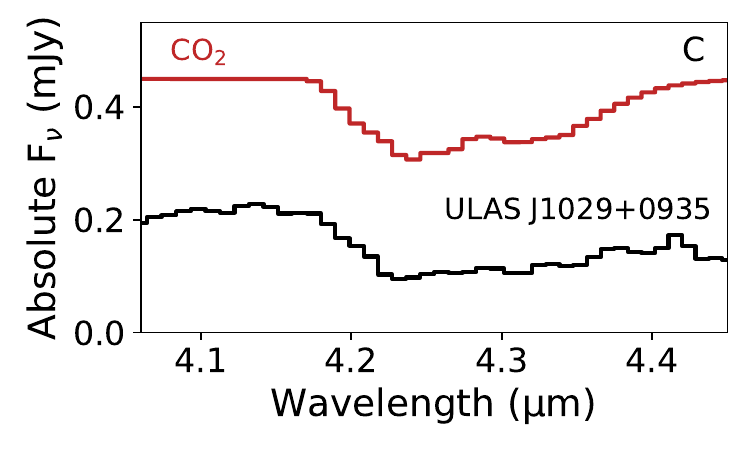} 
\\ \\
\includegraphics[width=0.9\textwidth]{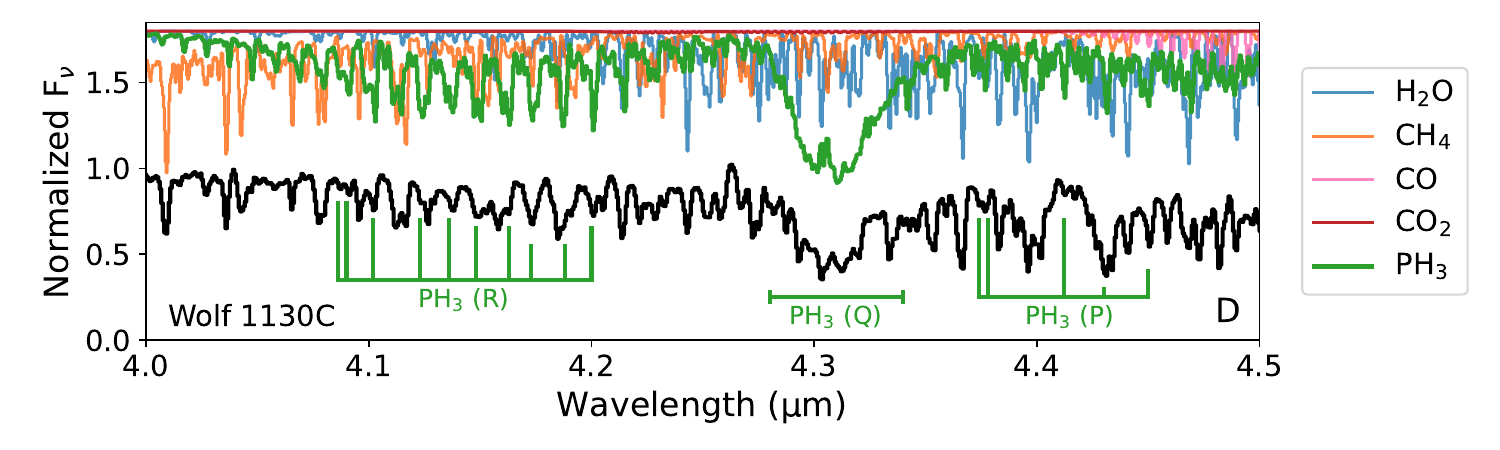} \\
\end{minipage}
\caption{{\bf JWST NIRSpec spectra of {\name} compared to ULAS~J1029+0935 and models.}}
{(A)}
The 0.9 to 5.2~{\micron} {NIRSpec/Prism} spectra of 
{\name} (black solid line) and 
%Ross~458C (magenta dashed line)
% ULAS~J102940.52+093514.6
the brown dwarf ULAS~J1029+0935 (magenta dashed line \cite{2024ApJ...973..107B})
in absolute $F_\nu$ flux densities (flux per unit frequency) measured in milliJanskies (mJy).
Colored bars indicate identified molecular absorption features in this region, as labeled.
We also indicate the 1.4 to 2.8 {\micron} region where collision-induced absorption (CIA) from {\hh} influences low-temperature spectra (horizontal black line).
The box centered at 4.25~{\micron} highlights the region shown in panels B and C.
%indicated; features not directly confirmed in this study are labeled with parentheses.
%JWST/NIRSpec spectra of {\name} (top) and Ross~458C (bottom) in the 3--5~{\micron} region. Primary molecular absorption features are labeled, and the region around coincident PH$_3$ and {\coo} bands at 4.2~{\micron} is highlighted. 
{(B) 
4.05 to 4.45~{\micron} region of the spectrum of {\name} (black line)
compared to a theoretical absorption spectrum of {\phhh} at an equivalent temperature (green line).
(C) 
Same as panel (B) but comparing the spectrum of ULAS~J1029+0935 (black line)
to a theoretical absorption spectrum of {\coo} (red line).
% {\phhh} \cite{10.1093/mnras/stu2246} and 
% {\coo} \cite{10.1093/mnras/staa1874} which clearly shape the respective spectra.}
%  and 
% {\coo} , which clearly reproduce the observations.}
(D)
Normalized NIRSpec/G395H spectrum of 
{\name} in the 4.0 to 4.5~{\micron} region (black line) compared to
%(middle panel) and Ross~458C (bottom panel) 
%Normalized spectra are shown as black lines, while color lines display 
theoretical absorption spectra for 
{\hho} (blue), 
{\chhhh} (orange), 
{\co} (pink), 
{\coo} (red), and 
{\phhh} (green) 
% {\hho} (blue \cite{2018MNRAS.480.2597P}), 
% {\chhhh} (orange \cite{10.1093/mnras/stae148}), 
% {\co} (pink \cite{2015ApJS..216...15L}), 
% {\coo} (red \cite{10.1093/mnras/staa1874}), and 
% {\phhh} (green \cite{10.1093/mnras/stu2246}) 
at the temperature and relative abundances {inferred from}
%for these atmospheres. For {\name}, these abundances are based on 
our retrieval analysis (Table~\ref{tab:fit_results}). 
Green lines and labels indicate the identified
blended Q-branch and individual R- and P-branch lines of {\phhh}.
}
\label{fig:spectra}
\end{figure}

The presence of {strong} phosphine absorption in the {spectrum} of
%of the cold and {metal-depleted} brown dwarf 
{\name} is unlike other observed low-temperature brown dwarf and giant exoplanet spectra, which are highly depleted in {\phhh} relative to 
vertical mixing models
\cite{2024ApJ...973...60B,2024ApJ...977L..49R}.
% As noted, {\phhh} should be abundant in these atmospheres given the evidence of vigorous vertical mixing as traced by other molecules (e.g., CO/CH$_4$, N$_2$/NH$_3$ \cite{1999ApJ...519L..85G,2006ApJ...647..552S}),
% and 
% Our inferred {\phhh} abundance for {\name}
% is quantitatively consistent with both equilibrium and vertical mixing models  (Fig.~\ref{fig:chemistry}), making it the first low-temperature atmosphere beyond Jupiter and Saturn for which phosphorus chemistry behaves as predicted.
The higher abundance of {\phhh} 
in the atmosphere of {\name} may be due to its low metallicity.
All other effects being equal, reducing the elemental abundances of carbon and oxygen by a factor of ten reduces the {\coo} abundance by a factor of one thousand (Fig.~\ref{fig:chemistry2}), which in turn reduces the strength of the prominent 4.3~{\micron} {\coo} $\nu_2$ absorption band. Removal of this band may explain why
%On its own, this depletion is expected to reduce {\coo} equilibrium abundances and corresponding absorption strength by a factor of $\sim$100 and ``unmasking'' 
the weaker 4.3~{\micron} {\phhh} feature can be observed in {\name} (Fig.~\ref{fig:spectra}). 

\begin{figure}[h]
\centering
%A\hspace{2.7in}B \\ \vspace{0.3cm}
 \includegraphics[width=0.42\textwidth]{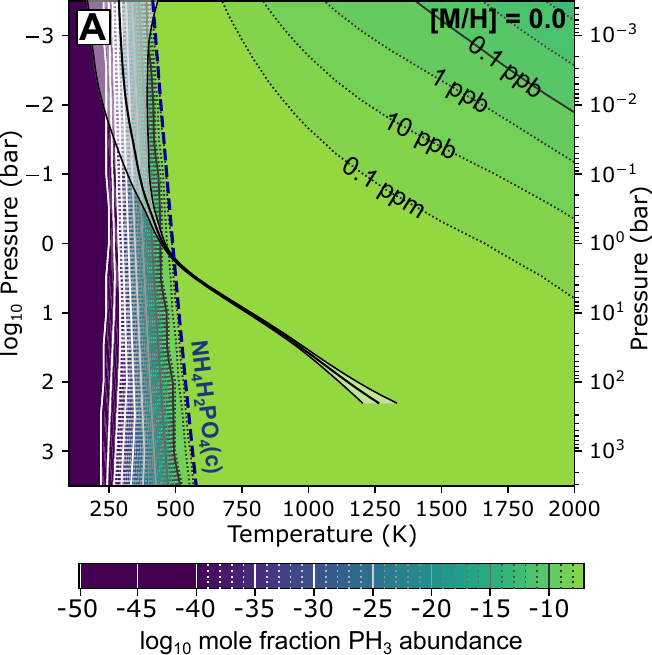} 
 \hspace{0.5cm} \includegraphics[width=0.42\textwidth]{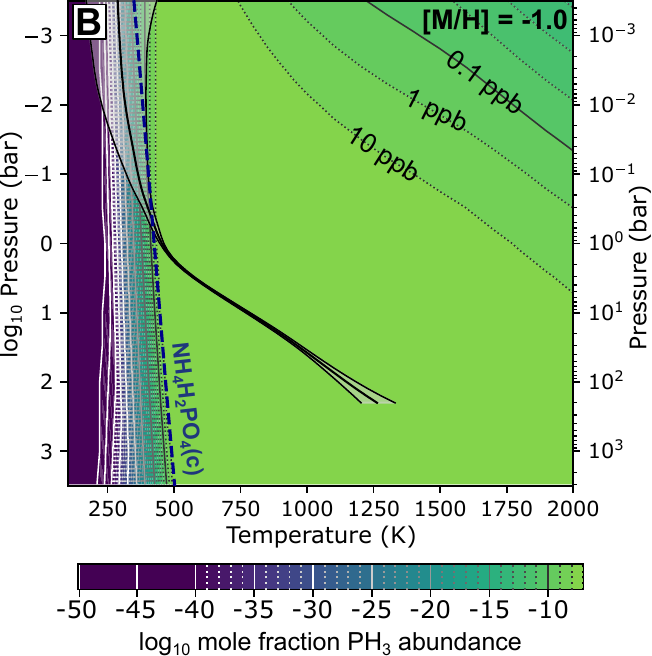} \\
 \vspace{1cm}
%C \\ \vspace{0.3cm}
\includegraphics[width=0.42\textwidth]{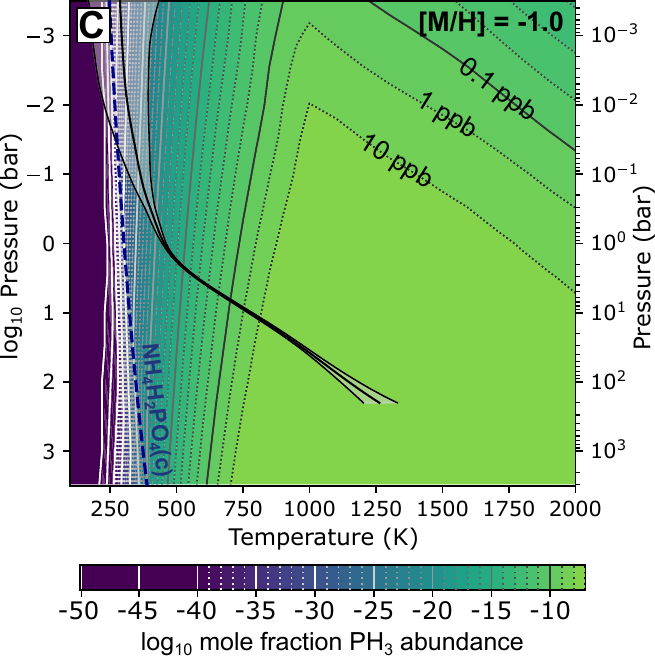} 
\caption{{\bf Chemical equilibrium model predictions of {\phhh} abundances in the atmosphere of {\name}.}
Each panel displays the equilibrium mole fraction abundances for {\phhh}
in green to purple shading as a function of temperature and pressure.
Abundance values are indicated by the associated color bars, and isoabundances are indicated in 1~dex (dotted) and 5~dex (solid) increments (black on green, white on purple), with values between 0.1~ppb (parts per billion) through 0.1~ppm (parts per million) labeled.
%The left panel assumes a gas mixture with 
%solar metal abundances, the center and right panels assume one-tenth solar %metal abundances. All models assume a bulk C/O element ratio of 0.3.
The near-vertical black dashed line indicates the condensation curve for {\adp}, which depletes {\phhh} at lower temperatures.
The three black curves indicate the median and $\pm$1$\sigma$ uncertainty boundaries for the atmospheric thermal profile of {\name} derived from our retrieval analysis
(Fig.~\ref{fig:retrieval2}).
(A) 
A model with solar metallicity and nominal phosphorus chemistry. 
(B) Same as panel A, but for one tenth of solar metallicity. 
(C) Same as panel B, but assuming a more exothermic enthalpy of formation for P$_4$O$_6$,
%\cite{chase1998}
leading to depletion of {\phhh} at warmer temperatures \cite{supp}.
}
\label{fig:chemistry}
\end{figure}

However, lower {\coo} opacity does not explain why the overall abundance of {\phhh} in {\name}
is consistent with vertical mixing models, while this molecule is highly depleted in other brown dwarfs and {gas} giant exoplanets.
%Variations in elemental abundances can influence the which chemical reactions dominate in a low-temperature atmosphere.
Our chemical equilibrium models (Fig.~\ref{fig:chemistry}) show that the formation of {\adp}, a condensate that removes {\phhh} gas from low-temperature atmospheres, occurs at a lower temperature and pressure in a more metal-poor atmosphere \cite{supp}. This shift in the condensation curve allows {\phhh} to exist over a wider portion of the photosphere compared to metal-enriched atmospheres. However, both Jupiter and Saturn {have} metal-enriched atmospheres and are considerably colder than {\name} ({\teff} $\approx$ 100~K compared to $\approx$ 600~K),
which is inconsistent with delayed condensation as an explanation.
We also consider the role of intermediate phosphorus oxides in the depletion of {\phhh} \cite{supp}, including P$_4$O$_6$ which has poorly constrained thermodynamic properties \cite{2020JGRE..12506526V}.
% DROPPED ,2023ESC.....7.1219B}.
We find that varying the chemical properties of P$_4$O$_6$ reduces {\phhh} equilibrium abundances, matching those of other brown dwarfs and giant exoplanets, but fails to explain the higher abundances in {\name}, Jupiter, and Saturn.
%cannot be simultaneously explained by such a change.
%argues against a universal thermochemical explanation. Moreover, 
%Equilibrium abundances are supplanted by vertical mixing that brings {\phhh} up from the deeper, warmer troposphere, a process that appears to be ubiquitous in brown dwarf and gas giant planet atmospheres.

{\name}'s high {\phhh} abundance could alternatively be due to
%inventory may emerge from its}
%The strength of the phosphine band in the spectrum of {\name} may also be relatively enhanced by the 
modest enrichment of elemental phosphorus
%in the atmosphere, 
inferred from our retrieval analysis. 
Previous studies of metal-poor stars have found phosphorus is enriched to levels exceeding the predictions of Galactic nucleosynthesis models \cite{2012A&A...540A..33C,2014ApJ...796L..24J}, by up to an order of magnitude
\cite{2020NatCo..11.3759M}.
One possibility is that the massive white dwarf component of this system, Wolf~1130B, could be a local source of phosphorus enhancement.
%This object is the remnant of a 6 to 8 solar mass progenitor star \cite{2018ApJ...854..145M}.
%in some cases well beyond predictions  \cite{2012A&A...540A..33C,2014ApJ...796L..24J}.
%While this enrichment likely arises through {broader Galactic evolutionary processes,} 
%a common processes that enrich phosphorus in other metal-poor stars, 
%we note that the
%enrichment in the atmosphere of {\name} important consideration is whether {\phhh} is uniquely enhanced in the atmosphere of {\name} due to an underlying enrichment of phosphorus in the 
%Notably, the} original 6--8 Solar mass ``primary'' of the Wolf~1130 system, now Wolf~1130B, could {be a local} phosphorus source.
The 6 to 8 solar mass progenitor of this component can produce phosphorus through slow neutron capture nucleosynthesis during a late giant phase of its evolution
%shell during a ``super''-asymptotic giant branch phase 
%\cite{2005NuPhA.758..259G,2011MNRAS.415.3865V,2020NatCo..11.3759M,2023Natur.623..292K 
\cite{2011MNRAS.415.3865V},
% %\cite{2008A&A...490..715P,2011ApJ...729...39O}
%
and enrich {\name} during the planetary nebular stage.
%\cite{2008A&A...490..715P} could have enriched {\name}. 
The current 1.2~{\msun} white dwarf remnant could in principle also
%that comprises the present remnant core of Wolf~1130B can continue to 
%, if an ONe type, 
%could also be a source of 
produce phosphorus through proton capture nucleosynthesis
of material deposited onto its surface by its close companion Wolf~1130A, and subsequently ejected during novae episodes 
%\cite{1998ApJ...494..680J,2014MNRAS.442.2058D,2024ApJ...967L...1B}.
\cite{2014MNRAS.442.2058D,2024ApJ...967L...1B}.
However, the spectrum of Wolf~1130A shows no evidence of phosphorus enrichment \cite{supp}. 
%in this much closer companion to Wolf~1130B, 
We therefore suggest that the slightly elevated abundance of phosphorus in {\name} is more likely to reflect the composition of metal-poor stars in the Milky Way's thick disk, rather than local production. 
%Phosphorus is primarily synthesized in the interiors of massive ($>$10~{\msun}) stars; however, intermediate-mass stars can efficiently produce phosphorus in a carbon-burning shell during the ``super''-asymptotic giant branch phase \cite{2005NuPhA.758..259G,2011MNRAS.415.3865V,2020NatCo..11.3759M,2023Natur.623..292K}.

We conclude that phosphine is present in {\name} at an abundance consistent with vertical mixing processes, similar to Jupiter and Saturn but unlike other brown dwarf and gas giant atmospheres. 
The inability of models to consistently explain all these sources indicates an incomplete understanding of phosphorus chemistry in low-temperature atmospheres. We therefore caution against the use of phosphine as a biosignature until these discrepant findings are resolved.
% On the other hand, this result can address challenges in our understanding of phosphorus nucleosynthesis and enrichment. The broad {\phhh} feature at 4.3~{\micron} is clearly seen in our low-resolution NIRSpec/PRISM spectrum of {\name}. Comparable data are now being acquired for low-temperature brown dwarfs in deep JWST spectral surveys out to kiloparsec distances,
% where the Milky Way's thick disk and halo populations dominate \cite{2024ApJ...962..177B}. 
% The 3--5~{\micron} spectra of these distant and long-lived brown dwarfs provide a new resource for studying phosphorus enrichment over the history of the Milky Way.

\clearpage

%%%%%%%%%%%%%%%% MAIN TEXT FIGURES %%%%%%%%%%%%%%%

%%%%%%%%%%%%%%%% MAIN TEXT TABLES %%%%%%%%%%%%%%%

% \begin{table} % Do NOT use \begin{table*}
% 	\centering
% 	% Captions go above tables
% 	\caption{\textbf{All captions must start with a short bold sentence, acting as a title.}
% 		Then explain what is being listed in the table, the meaning of each column etc.
% 		Captions are placed above tables.}
% 	\label{tab:example} % give each table a logical label name
	
% 	\begin{tabular}{lccc} % four columns, alignment for each
% 		\\
% 		\hline
% 		Sample & $A$ & $B$ & $C$\\
% 		 & (unit) & (unit) & (unit)\\
% 		\hline
% 		First & 1 & 2 & 3\\
% 		Second & 4 & 6 & 8\\
% 		Third & 5 & 7 & 9\\
% 		\hline
% 	\end{tabular}
% \end{table}

%%%%%%%%%%%%%%%% REFERENCES %%%%%%%%%%%%%%%

\clearpage % Clear all remaining figures and tables then start a new page

% The list of references goes after the main text and before the acknowledgements
% When preparing an initial submission, we recommend you use BibTeX, like this:
%
\bibliography{jwstph3} % for a file named science_template.bib
\bibliographystyle{sciencemag}

% After the paper has completed peer review and been revised ready for acceptance,
% you should comment out the lines above and copy-paste the contents of your .bbl
% file here instead. This will help ensure that our conversion software works correctly.
% Remember to re-run BibTeX first - check the timestamp!
%
% Example of the first three entries copy-pasted from science_template.bbl:
%
%\begin{thebibliography}{1}
%
%\bibitem{example}
%A.~N. {Author}, An example reference. \emph{Journal of Improbable Research}
%  \textbf{1}, 67 (2020).
%
%\bibitem{example2}
%F.~M. {Surname}, S.~{Author}, A second example. \emph{Interesting Research
%  Letters} \textbf{32}, 897 (2019).
%
%\bibitem{example_preprint}
%P.~{One}, P.~{Two}, P.~{Three}, {An unpublished preprint}. \emph{preprint}
%  (2021), arXiv:2101.12345.
%
%\end{thebibliography}

%%%%%%%%%%%%%%%% ACKNOWLEDGEMENTS %%%%%%%%%%%%%%%

\section*{Acknowledgments}
% Here you can thank helpful colleagues who did not meet the journal's authorship criteria, or
% provide other acknowledgements that don't fit the (compulsory) subheadings below.
% Formatting requirements for each of these sections differ between the \textit{Science}-family
% journals; consult the instructions to authors on the journal website for full details.

%\paragraph*{Acknowledgments:}
We thank our three anonymous referees for constructive criticism, which greatly improved this article. 
This work is based on observations made with the NASA/ESA/CSA James Webb Space Telescope. The data were obtained from the Mikulski Archive for Space Telescopes (MAST) operated by the Space Telescope Science Institute (STScI), which is operated by the Association of Universities for Research in Astronomy, Inc., under NASA contract NAS 5-03127 for JWST. These observations are associated with program \# GO-4668.
We thank Andrew Fox, Greg Sloan, Alaina Henry, and Wilson Joy Skipper at STScI for their assistance in the planning and execution of JWST observations.
This publication makes use of data products from the Wide-field Infrared Survey Explorer (WISE), which is a joint project of the University of California, Los Angeles, and the Jet Propulsion Laboratory/California Institute of Technology, funded by NASA.
Data from the European Space Agency (ESA) mission Gaia (\url{https://www.cosmos.esa.int/gaia}) were processed by the Gaia Data Processing and Analysis Consortium (DPAC; \url{https://www.cosmos.esa.int/web/gaia/dpac/consortium}). Funding for the DPAC has been provided by national institutions, in particular the institutions participating in the Gaia Multilateral Agreement. 
This research has made use of the SIMBAD database, operated at CDS, Strasbourg, France \cite{2000A&AS..143....9W}.

\paragraph*{Funding:}
% List the grants, fellowships etc. that funded the research;
% use initials to specify which author(s) were supported by each source.
% Include grant numbers when appropriate or required by the funding agency.
% For example: F.~A. was funded by the Generous Science Agency grant~2372.
AB, EG, C-CH, and GS acknowledge funding support from NASA/STScI through JWST general observer program GO-4668, under NASA contract NAS 5-03127.
AB, EG and CS-S acknowledge funding support from the Heising-Simons Foundation.
CS-S acknowledges funding support from the Portuguese Foundation for Science and Technology. 
BB acknowledges funding support from UK Research and Innovation Science and Technology Facilities Council, grant number ST/X001091/1.  
CV acknowledges funding support from NASA/STScI through JWST archival theory program AR-2232.

\paragraph*{Author contributions:}
% List each author’s contributions to the paper.
% Use initials to abbreviate author names.
AB was principal investigator of the JWST observing program, oversaw execution of the observations, conducted atmosphere grid modeling, and led writing of the manuscript.
EG led the atmospheric retrieval analysis with contributions from BB and JF.
SB contributed to the atmosphere grid modeling.
CV led the atmospheric chemistry modeling.
%GM provided spectra of Wolf~1130A and contributed to system characterization.
GM and ZZ contributed to analysis of the Wolf~1130 system trinary system.
%ZZ contributed to analysis of the metallicity of Wolf~1130A.
CS-S discussed the phosphine opacity and use as a biosignature.
NL, SM, RG, AM, MC, AS, GS, CT, CH, and CA contributed to preparing the JWST proposal and the manuscript writing.

\paragraph*{Competing interests:}
There are no competing interests to declare.

\paragraph*{Data and materials availability:}
% Specify where the data, software, physical samples, simulation outputs or other materials
% underlying the paper are archived.
% They must be publicly accessible when the paper is published (without embargo) and enable
% readers to reproduce all the results in the paper.
% Contact the editor if you’re unsure what needs to be shared.

% Our preference is for digital material to be deposited in a suitable non-profit online data or
% software repository that issues the material with a DOI.
% Alternatively, an institutional repository, subject-based archive, commercial repository etc.
% is acceptable, as are (short) supplementary tables or a machine-readable supplementary data file.
% ‘Available on request’ or personal web pages are not allowed.
The JWST data presented are 
%in this paper were obtained as part of program JWST-GO-04668 (PI Burgasser) 
%and GO-01277 (PI Lagage), 
%and are publicly 
available at the Mikulski Archive for Space Telescopes %(\url{https://archive.stsci.edu})
(\url{https://mast.stsci.edu/portal/Mashup/Clients/Mast/Portal.html})
under program JWST-GO-4668 for target name Wolf1130B, and
{as data collection DOI: 10.17909/5c42-wc87 \cite{jwstdata}.}
%Current and past versions of the JWST data reduction pipeline {\tt jwst} can be obtained at \url{https://jwst-pipeline.readthedocs.io/en/latest/}. 
The modified Sonora Elf-Owl models (EOwl+ \cite{2024ApJ...973...60B}) presented in Figure~\ref{fig:gridfit} are available at 
DOI:10.5281/zenodo.11370829 \cite{beiler_2024_11370830}.
\textsc{Brewster} code is available at 
The \textsc{Brewster} code is available at \url{https://github.com/substellar/brewster}
and archived on Zenodo \cite{gonzales_2025_17082357}.  
The \textsc{ucdmcmc} code is available at \url{https://github.com/aburgasser/ucdmcmc}
and archived on Zenodo \cite{ucdmcmc}.
Our best-fitting parameters models are listed in Table~\ref{tab:fit_results}.
%The {\tt emcee} code is publicly available at \url{https://github.com/dfm/emcee}.  
%Correspondence and requests for other materials should be addressed to Adam Burgasser (aburgasser@ucsd.edu). 

%%%%%%%%%%%%%%%% SUPPLEMENT LIST %%%%%%%%%%%%%%%

% List the contents of your Supplementary Materials, including the numbers of any
% supplementary figures, tables, external data files etc. and any references that are
% cited only in the supplement. In this example, refs. 7-8 are cited only in the supplement.
% Fill out your numbers accordingly and delete any lines that aren't applicable.
\subsection*{Supplementary Materials}
Materials and Methods\\
Supplementary Text\\
Tables S1 to {S2}\\
Figures S1 to {S6}\\
References \textit{(56-\arabic{enumiv})}\\ % automatically fills out the last reference number
% (filling out the other numbers automatically is possible but fiddly and liable to break)
%Movie S1\\
%Data S1

%%%%%%%%%%%%%%%% END OF MAIN TEXT %%%%%%%%%%%%%%%

\newpage

%%%%%%%%%%%%%%%% START OF SUPPLEMENT %%%%%%%%%%%%%%%

% Figures, tables, equations and pages in the supplement are numbered S1, S2 etc.
\renewcommand{\thefigure}{S\arabic{figure}}
\renewcommand{\thetable}{S\arabic{table}}
\renewcommand{\theequation}{S\arabic{equation}}
\renewcommand{\thepage}{S\arabic{page}}
\setcounter{figure}{0}
\setcounter{table}{0}
\setcounter{equation}{0}
\setcounter{page}{1} % not 0 as \newpage already started a supplementary page
% References continue the numbering from the main text.

%%%%%%%%%%%%%%%% SUPPLEMENT TITLE PAGE %%%%%%%%%%%%%%%

\begin{center}
\section*{Supplementary Materials for\\ \scititle}

% Author list for the supplement
% Indicate the corresponding authors, but do NOT include institutions here
% It would be nice if the template auto-generated this, but doing so is complicated...
%%%
%%% REPEAT FROM ABOVE
%%%
Adam J.\ Burgasser$^{\ast}$,
Eileen C.\ Gonzales,
Samuel A.\ Beiler,
Channon Visscher, \\
Ben Burningham,
Gregory N.\ Mace,
Jacqueline K.\ Faherty,
Zenghua Zhang, \\
Clara Sousa-Silva,
%% proposing team CoPIS
Nicolas Lodieu,
Stanimir A.\ Metchev,
%% proposing team CoIS
Aaron Meisner, \\
Michael Cushing,
Adam C.\ Schneider,
Genaro Suarez,
Chih-Chun Hsu, \\
Roman Gerasimov,
Christian Aganze,
and Christopher A. Theissen\\
\small$^\ast$Corresponding author. Email: aburgasser@ucsd.edu\\
\end{center}

% Fill out the numbers for each type of supplementary material,
% and delete any lines that aren't applicable.
% These are just example numbers that don't match the rest of this template.
\subsubsection*{This PDF file includes:}
Materials and Methods\\
Supplementary Text\\
Figures S1 to {S6}\\
Tables S1 to {S2}\\
%References \textit{(51-\arabic{enumiv})}\\

%\subsubsection*{Other Supplementary Materials for this manuscript:}
%Movies S1 to S2\\
%Data S1 to S2

\newpage

%%%%%%%%%%%%%%%% MATERIALS AND METHODS %%%%%%%%%%%%%%%

% The Materials and Methods section should contain details of the samples measured,
% experiments performed, observations taken, simulations run, data analysis, statistical methods etc.
% Give enough detail for any competent researcher in your field to fully reproduce the results.

% To refer to this section from the main text, use the numbered note in the reference list \cite{methods}.
% Refer to figures and tables in the same way as in the main text but now all capitalized e.g.
% Fig.~\ref{fig:example}, Table~\ref{tab:example},
% Fig.~\ref{fig:sup_example} and Table~\ref{tab:sup_example}.
% Cite references in the usual way \cite{example2},
% including any that are only cited in the supplement \cite{sm_example,conference_example}.

\subsection*{Materials and Methods}

\subsubsection*{JWST/NIRSpec Observations}
{\name} was observed with JWST/NIRSpec and JWST/MIRI on 2024 Aug 30 as part of program JWST-GO-4668 {\cite{jwstdata}}.  
NIRSpec/Prism data were acquired using the S200A1 0$\farcs$2 fixed slit and Clear filter to obtain low-resolution spectra (resolving power {\ldl} = 50--300) spanning 0.6 to 5.2~{\micron}.  
{\name} was acquired with the F110W filter
and offset to the slit position, then observed in spectral mode at two spatial nod positions along the slit, with a sub-pixel dither by 0.5 pixels, equivalent to 0$\farcs$05. At each nod position we obtained two exposures
consisting of 160 sample-up-the-ramp groups with 1.6~{seconds} integrations per group, for a total exposure time of 997~{seconds}. 

NIRSpec/G395H data were acquired using the S200A1 plus S200A2 slit combination and the F290LP filter to obtain moderate-resolution spectra ({\ldl} = 2,000--3,700) spanning 2.9 to 5.1~{\micron}.
The dual slit {enabled} coverage of the 3.7 to 3.8~{\micron} gap between NIRSpec's two detectors.
{\name} was observed in two spatial nod positions for each slit,
with 50 sample-up-the-ramp groups and 5.5~{seconds} integrations per group, for a total exposure time of 1099~{seconds}.

MIRI/LRS data were acquired using the 4$\farcs$7$\times$0$\farcs$51 (42.7$\times$4.6~pixel) slit to obtain low-resolution spectra ({\ldl} = 40--160) spanning 4 to 14~{\micron}. 
{\name} was acquired with the F1000W filter
and offset to the slit position, then observed in spectral mode at two spatial nod positions along the slit separated by 1$\farcs$9. At each slit position, one exposure was obtained consisting of 100 sample-up-the-ramp groups with 2.8~{seconds} integrations per group and the FASTR1 readout pattern, for a total exposure time of 555~{seconds}. 

Because {\name} was well-centered in all spectral exposures, we analyzed the standard pipeline data reduction products.
%served through the Barbara A.\ Mikulski Archive for Space Telescopes (MAST). 
The JWST science calibration pipeline version 1.15.1 \cite{2024zndo..12692459B} was used for all data, which applied background, dark current, and bias subtraction; flat field, gain scale, and linearity corrections; bad pixel flagging; count rate extraction; saturation correction; and 1D spectral extraction. 
The final one-dimensional spectra have median signal-to-noise (S/N) per pixel values of 300 for the NIRSpec/Prism data and 75 for the NIRSppec/G395H data at 4.1~{\micron}, and S/N = 90 for the MIRI/LRS data at 9~{\micron}.
%{These data can be accessed through MAST collection DOI:10.17909/5c42-wc87 \cite{jwstdata}.}

\subsubsection*{Luminosity and Effective Temperature of {\name}\label{sec:sedfit}}

The infrared spectrum of {\name} includes a large proportion of its total flux, allowing us to estimate its bolometric luminosity.  We followed previous work \cite{2024ApJ...973..107B} to build the spectral energy distribution (SED) by combining the NIRSpec/Prism and NIRSPec/MIRI spectra.
% with 4 to 14~{\micron} low resolving-power spectra ({\ldl} $\approx$ 100) obtained with the 
% the JWST Mid-Infrared Imager (MIRI \cite{2023PASP..135d8003W}),
% also through program JWST-GO-04668. Details on these observations will be presented in a forthcoming publication. 
{The combined spectrum was calibrated to apparent flux densities using a Wide Field Infrared Explorer (WISE \cite{2010AJ....140.1868W}) W2 (4.6~{\micron}) magnitude of {\name} of 15.14$\pm$0.02 (Vega mag)
based on photometry from the CatWISE2020 catalog \cite{2021ApJS..253....8M,2025A&A...698A.141Z}.}
We extrapolated the first spectral point at $\lambda$ = 0.55~{\micron} to zero flux at $\lambda$ = 0, and extended a Rayleigh-Jeans tail from the last spectral point at $\lambda$ = 14~{\micron} to $\lambda$ = $\infty$. Integrating over this SED yields a bolometric flux $F_\mathrm{bol} = (9.940\pm0.006)\times10^{-17}$~W~m$^{-2}$. 
The parallax of the primary star Wolf 1130A from Gaia data release 3 \cite{2023A&A...674A...1G}, 60.30$\pm$0.03~mas,  determines the distance to the system, which we adopt for {\name}. This measure yields a bolometric luminosity relative to the Sun of {\logl} = $-$6.068$\pm$0.002. 
Because there are absorption features redward of 14~{\micron} in the spectra of low-temperature brown dwarfs, 
%it is necessary to 
{we used} a previously computed correction to the Rayleigh-Jeans approximation {determined}
from a sample of brown dwarfs
%we determined a linear correction to $L_\mathrm{bol}$ 
as a function of absolute W2 magnitude,
\begin{equation}
    \Delta\log(L_\mathrm{bol}) = 0.0238{\times}M_\mathrm{W2} - 0.3133 
\end{equation}
\cite{2024ApJ...973..107B}, which has an uncertainty of 0.010~dex at $M_\mathrm{W2}$ = 14.04 mag.
Applying this correction yields {\logl} = $-$6.047$\pm$0.003. 
The age of this system is currently unconstrained, so we assume 8 to 15 Gyr, which is typical for estimates of the Milky Way's thick disk population \cite{2017ApJ...837..162K,2019MNRAS.490.5335S}.
Using these values as input for the
%Combining this luminosity with an {estimated} age of 8$-$15 Gyr based on the system's thick disk membership, the 
Sonora Bobcat evolutionary models \cite{2021ApJ...920...85M} for a solar-scaled metallicity of [M/H] = $-$0.5 leads to the prediction of an effective temperature of {\teff} = $621\pm9$~K and a radius R = $0.80\pm0.02$ {Jupiter radii ({\rjup})}, values that are used to verify our atmosphere model fits.

\subsubsection*{Spectral Model Fits to Prism Data\label{sec:gridfit}}

We fit models to
%conducted an initial assessment of the atmosphere parameters of {\name} by fitting 
the absolute flux-calibrated NIRSpec/Prism spectrum using 
two versions of the Sonora Elf-Owl atmosphere models: 
the nominal version in which {\phhh} is artificially suppressed to account for its absence in observed brown dwarf spectra (hereafter EOwl \cite{2024ApJ...963...73M}) 
and an expanded chemistry version in which
{\phhh} abundances are determined from vertical mixing (hereafter EOwl+ \cite{2024ApJ...973...60B}).
For both versions we considered models with
effective temperatures 400~K~$\leq$~{\teff}~$\leq$~1000~K,
log surface gravities 4.0~$\leq$~$\log_{10}\left(\mathrm{g/cm~s^{-2}}\right)$~$\leq$~5.5,
log solar-scaled metal abundances $-$1~$\leq$~[M/H]~$\leq$~1, 
% ([M/H]$_\odot$ = 0),
C/O linear abundance ratios 0.22 $\leq$ C/O $\leq$ 1.14 (as compared to a solar C/O abundance ratio of 0.458 \cite{2003ApJ...591.1220L})
and log vertical mixing diffusion coefficients 
2~$\leq$~{\kzz}~$\leq$~9.
Model fitting was conducted 
using the \textsc{ucdmcmc} package version 1.1, %\footnote{\url{https://github.com/aburgasser/ucdmcmc}}, 
which is archived on Zenodo \cite{ucdmcmc}. 
%The models used and their parameter ranges are summarized in Table~\ref{tab:models}.
The observed spectrum 
was calibrated to absolute flux densities $F_\lambda$ (flux per unit wavelength) 
using an absolute W2 magnitude of 14.04$\pm$0.02 \cite{2021ApJS..253....8M,2023A&A...674A...1G,2025A&A...698A.141Z}. The
observed spectrum and models were interpolated onto a common wavelength scale
that accounts for the variable resolution of the NIRSpec/Prism dispersion, and
the models were additionally smoothed by  2~pixel-wide Gaussian profile to account for the projected slit width on the detector. 
%The observed spectrum was scaled to absolute flux densities  on a $F_\lambda$ scale using an absolute W2 (4.6~{\micron}) magnitude of 14.04$\pm$0.02 (Vega mag) based on photometry from the CATWISE-2020 catalog \cite{2021ApJS..253....8M} and the Gaia parallax of Wolf~1130A \cite{2023A&A...674A...1G}.

We conducted an initial grid search to find the single best-fitting model {in each of the EOwl} model sets,
{then used these parameters to seed a Metropolis-Hastings Markov Chain Monte Carlo (MCMC) fitting algorithm \cite{1953JChPh..21.1087M,HASTINGS01041970}. Models were linearly interpolated across the parameter grid using the logarithm of flux densities and logarithmic parameter values.
%({\teff} $\rightarrow$ $\log${\teff}). 
We used a single chain of 5,000 steps and a reduced $\chi^2_r$ statistic to iterate model parameters:}
\begin{equation}\label{eqn:chi}
    \chi^2_r = \frac{1}{DOF}\sum\limits_{i=1}^N\frac{(O_i-\alpha{M_i})^2}{\sigma_i^2}.
\end{equation}
Here, $DOF$ is the degrees of freedom, equal to the number of spectral data points (N = 820) minus six free fitting parameters;
$O_i$ is the observed spectral flux density; 
$\sigma_i$ is the observed {spectral} uncertainties;
$M_i$ is the model spectrum;
and $\alpha$ is a scaling factor that minimizes $\chi^2_r$ {\cite{2008ApJ...678.1372C}}: 
\begin{equation}\label{eqn:alpha}
    \alpha=\frac{\sum\limits_{i=1}^NM_iO_i/\sigma_i^2}{\sum\limits_{i=1}^N{M_i^2/\sigma_i^2}}.
\end{equation}
The sum was carried out over all spectral bins $i$ that satisfied $O_i/\sigma_i > 2$. Because the models were computed in surface fluxes and the observed spectrum is scaled to absolute fluxes, corresponding to a distance $d$ = 10~pc, the optimal scale factor {$\alpha = \left({R}/10~\mathrm{pc}\right)^2$} provides an estimate of the source radius, $R$ = {4.316$\times$10$^{9}\sqrt\alpha$~{\rjup}}.
Further details of the fitting procedure are described elsewhere \cite{2025ApJ...982...79B}.

\begin{figure}[h]
\centering
\includegraphics[width=0.9\textwidth]{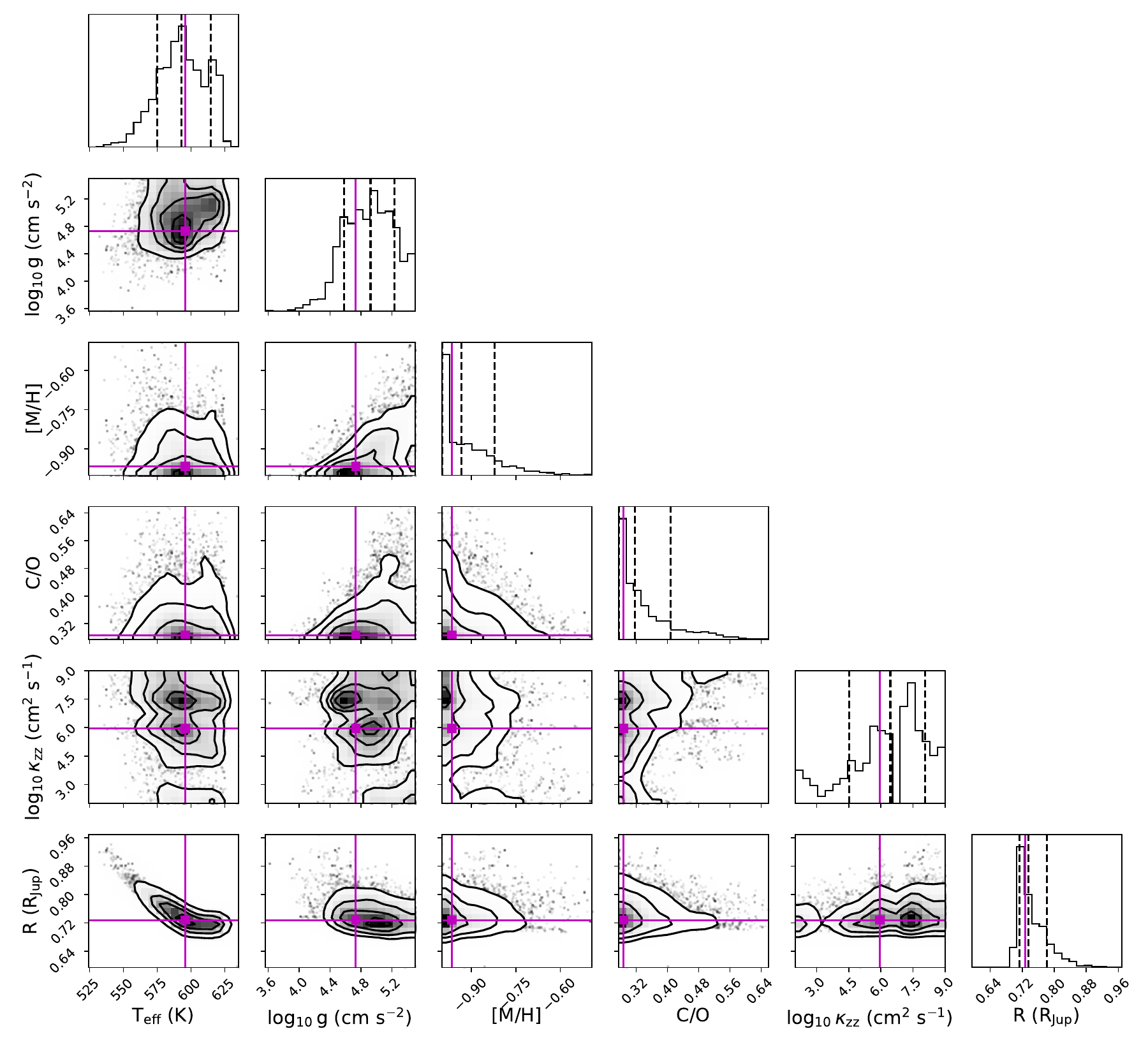} 
\caption{{\bf Posterior probability distributions for Sonora Elf-Owl models fitted to the  spectrum of {\name}}. 
Results are shown for fits based on models with {\phhh} abundances set by vertical mixing (EOwl+ \cite{2024ApJ...973...60B}).
Panels along the diagonal show histograms of the marginalized posterior distributions for model parameters {\teff}, {\logg}, [M/H], C/O, $\log\kappa_{zz}$, and radius $R$. 
Off-diagonal panels display the marginalized posterior distributions among parameter pairs as contour plots, with contours set at 25\%, 50\% and 75\% confidence intervals. 
%indicating correlations between parameters.
Solid magenta lines in the posterior distributions indicate the best-fitting value of each parameter, while dashed black lines in the diagonal panels indicate quantiles of 16\%, 50\% and 84\% of the posterior distributions, listed in Table~\ref{tab:fit_results}.
This figure was generated using the \textsc{corner.py} package \cite{corner}.
}
\label{fig:gridfit-corner}
\end{figure}

Fig.~\ref{fig:gridfit} displays the best-fitting models for the EOwl and EOwl+ grids, while Table~\ref{tab:fit_results} and Fig.~\ref{fig:gridfit-corner} present the parameters and uncertainties from the EOwl+ grid fit derived from the MCMC posterior distributions.
Both versions of the EOwl models reproduce the overall SED of {\name} across the 1--5~{\micron} band, with $\sim$20\% deviations at the 1.0~{\micron} peak shaped by strong {\hho} and {\chhhh} bands, and at the 3~{\micron} peak where the model predicts excess absorption from {\nhhh}.
Both models also produce equivalent physical parameters, including a common radius of $R$ = 0.74$^{+0.04}_{-0.02}$~{\rjup}. These parameters are consistent with theoretical expectations for an 
old, evolved brown dwarf \cite{2021ApJ...920...85M}.
However, the EOwl+ models provide a better fit to the 4.0 to 4.5~{\micron} region encompassing the observed {\phhh} feature. This region is reproduced by 
assuming vertical mixing of {\kzz} = 6.5$^{+1.6}_{-1.9}$.
Using an F-test statistic, we find there is no significant difference between the best EOwl and EOwl+ models over the wavelength range 0.8 to 4.0~{\micron} (p-value = 0.15), but a significant improvement in the fit (p-value $< 0.001$) for the EOwl+ model over the wavelength range 4.0 to 4.4~{\micron} encompassing the {\phhh} feature. Based on this analysis, we conclude that {\phhh} is present at an abundance that is consistent with predictions from vertical mixing chemistry.

\subsubsection*{Phosphine and Carbon Dioxide Abundance\label{sec:abundfit}}

The abundances of {\phhh} and {\coo} in the atmosphere of {\name} 
were determined using the \textsc{PICASO} radiative transfer code \cite{2023ApJ...942...71M}, following previous work
\cite{2024ApJ...973...60B}. 
We started from the EOwl+ model grid,
with models encompassing 
effective temperatures 575~K~$\leq$~{\teff}~$\leq$~750~K;
log surface gravities 3.25~$\leq$~$\log_{10}\left(\mathrm{g/cm~s^{-2}}\right)$~$\leq$~5.5;
log solar-scaled metallicities [M/H] = $-$1 and 0 dex; 
C/O linear abundance ratios C/O = 0.5, 1.0, and 2.5 relative to solar; and 
log vertical mixing diffusion coefficients {\kzz} = 2, 4, 7, and 9.
Using the same pressure-temperature profile as the EOwl+ model and holding all other abundances constant, we explored enhancements in the photospheric abundances of 
{\phhh} (45~ppb $\times$0.2, $\times$0.5, $\times$1, and $\times$2) and {\coo} (9.4$\times$10$^{-5}$~ppb $\times$1, $\times$100, $\times$500, $\times$1,000, $\times$5,000, and $\times$10,000),
and compared our grid of models to the NIRSpec/Prism spectrum of {\name} as above.

%Sec.~\ref{sec:gridfit}.}
%, smoothed to account for the 
%variable resolving power of the data and two-pixel slit width \cite{2023ApJ...951L..48B},
%and uses the same $\chi^2$ statistic as Eqn.~\ref{eqn:chi}.

\begin{figure}[h]
\centering
\includegraphics[width=0.8\textwidth]{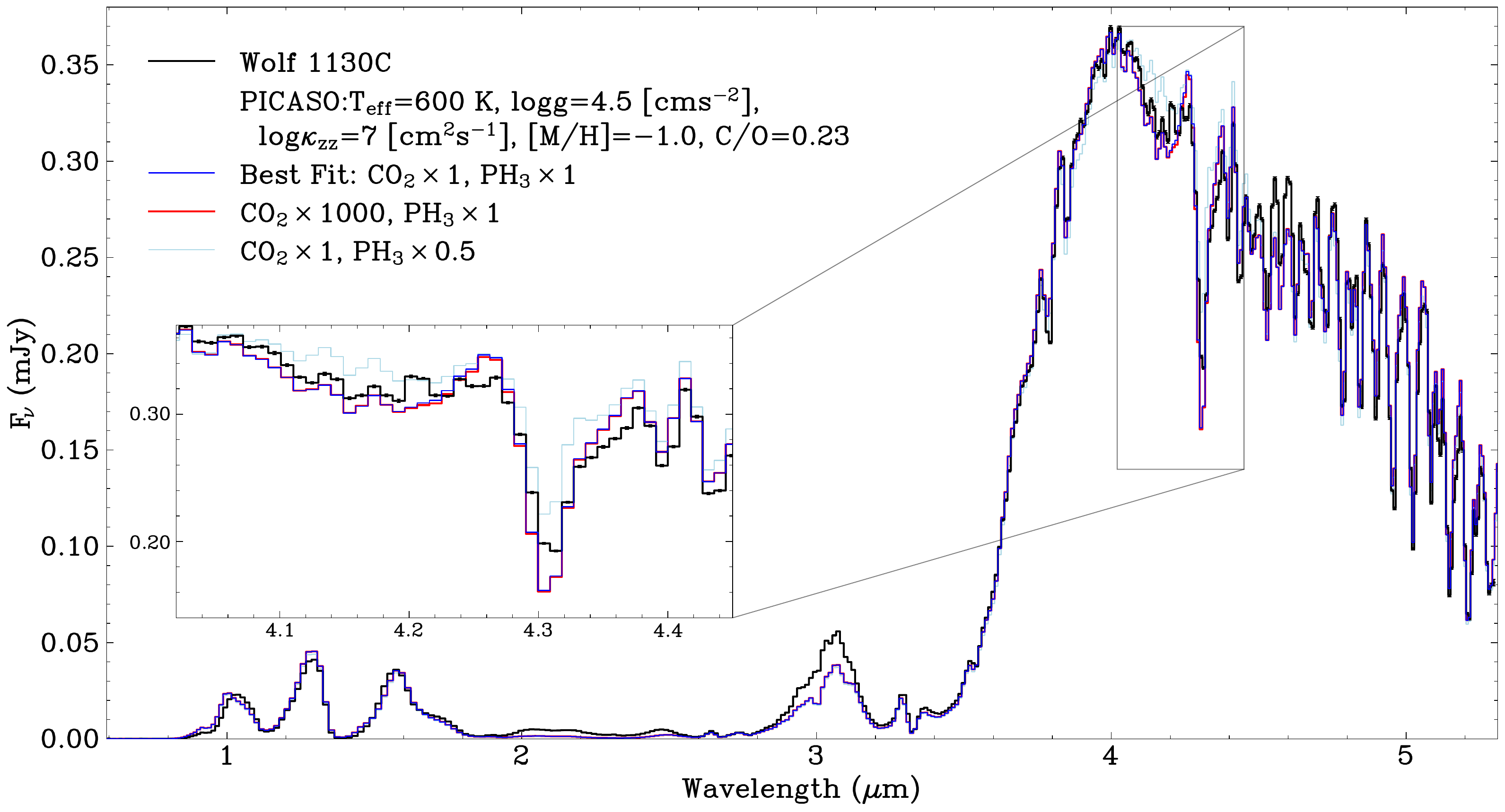}
\caption{{\bf PH$_3$ and CO$_2$ abundance analysis for {\name}.}
The NIRSpec/Prism spectrum of {\name} (black line \cite{jwstdata}) is compared to {three} PICASO models with 
{\teff}~=~600~K, 
$\log_{10}\left(\mathrm{g/cm~s^{-2}}\right)$~=~4.5, 
{\kzz}~=~7.
[M/H]~=~$-$1.0, and
C/O~=~0.23.
The blue line shows the best-fitting model with vertical mixing; 
the {nearly identical} red line 
%(which is nearly identical to the blue line) 
shows the best-fitting model with 1,000$\times$ enhancement in {\coo} and no change in {\phhh}; 
the light blue line shows {0.5$\times$ reduction} in {\phhh} and no change in {\coo}. 
The inset box zooms into the region around the 4.1~{\micron} {\phhh} feature to illustrate the differences between those models.
%and allow us to rule out significant reduction in {\phhh} and significant enhancement ($\geq$1,000$\times$) in CO$_2$.
%, and shows  close-up view, and has marginal significance.
}
\label{fig:picaso}
\end{figure}

For a solar-scaled abundance of [M/H] = $-$1, we find a best-fitting model of {\teff} = 600~K, {\logg} = 4.5, $\log_{10}\kappa_{zz}$ = 7, and C/O = 0.23 (half solar), with no enhancement in PH$_3$ or {\coo} (Fig.~\ref{fig:picaso}). 
These physical parameters are consistent with those inferred from our grid model fitting.
The differences between {\coo} enhancement values were minimal due to the absence of a clear feature,
although we rule out models with enhancements 500$\times$ or larger based on comparison of $\chi^2$ values through the Bayesian Information Criterion (BIC; \cite{10.1214/aos/1176344136}), which yields $\Delta{BIC}$ $\ge$ 26 relative to the {model with nominal abundances}, indicating very strong evidence against this level of enhancement \cite{kass1995}.
Similarly, a change in {\phhh} to lower and higher values yields 
%$\Delta\chi^2_r$ = 11.53 (
$\Delta$BIC $\ge$ 4,718 and 81,323, respectively, indicating very strong evidence against these variations. 
From this analysis, we conclude that {both} {\phhh} and {\coo} abundances in {\name} are consistent with predictions of vertical mixing models.
%, contrary to the findings for solar-metallicity late T and Y dwarfs that require {\phhh} depletion %and significant {\coo} enrichment \cite{2024ApJ...973...60B}.

\subsubsection*{Retrieval Modeling\label{sec:retrieval}}

We conducted {an atmospheric} retrieval analysis of the absolute flux-calibrated NIRSpec/G395H spectrum of {\name} using the \textsc{Brewster} package
\cite{2017MNRAS.470.1177B,2021MNRAS.506.1944B,2021ApJ...923...19G} following previous methods \cite{2024Natur.628..511F} 
with the following modifications \cite{gonzales_2025_17082357}.
\textsc{Brewster} combines an atmospheric forward model generator with a Bayesian sampler (\textsc{emcee} \cite{emcee}) to infer the best-fitting pressure-temperature (P/T) profile, molecular gas abundances ($f_\mathrm{i}$), global atmosphere physical parameters ({\logg}, radius), and kinematic parameters (radial and rotational velocities) that reproduce an observed spectrum. 
The assumed priors for these parameters are summarized in Table~\ref{tab:priors}.  
The gas molecules considered were 
{\hho} \cite{2018MNRAS.480.2597P}, 
{\chhhh} \cite{2020ApJS..247...55H}, 
{\co} \cite{2010JQSRT.111.2139R,2015ApJS..216...15L}, 
{\coo} \cite{2014JQSRT.147..134H}, 
{\nhhh} \cite{2011MNRAS.413.1828Y,2012ApJ...750...74S}, 
{\hhs} \cite{10.1093/mnras/stw1133}, 
and {\phhh} \cite{10.1093/mnras/stu2246}, 
and additional opacities drawn from the literature \cite{2010JQSRT.111.2139R,2008ApJS..174..504F,2014ApJS..214...25F,2007A&A...465.1085A,2007EPJD...44..507A}.
% Burningham: Freedman x 2, Ryab..., Allard x 2, Richard, Saumon
% Faherty CH4: Hargreaves
% Rowland CO: Rothman, Li
Volume mixing ratios 
assumed to be vertically constant throughout the photosphere. 
We did not include cloud opacity in this analysis. %\cite{2017MNRAS.470.1177B,2021MNRAS.506.1944B,2024ApJ...972..172P}. 
We ran 110,000 iterations with 16 walkers per parameter.
% {We also tested two models to show the impact of {\nhhh} and {\hhs} with 35,000-65,000 iterations. In each model, we include all gases in the best fit model minus the either {\nhhh} or {\hhs}. Both gases are strongly preferred; the model without {\nhhh} had a $\Delta{BIC}$ = 212, and the model without {\hhs} had a $\Delta{BIC}$ = 6.}

\begin{figure}[h]
\centering
\includegraphics[width=0.95\textwidth]{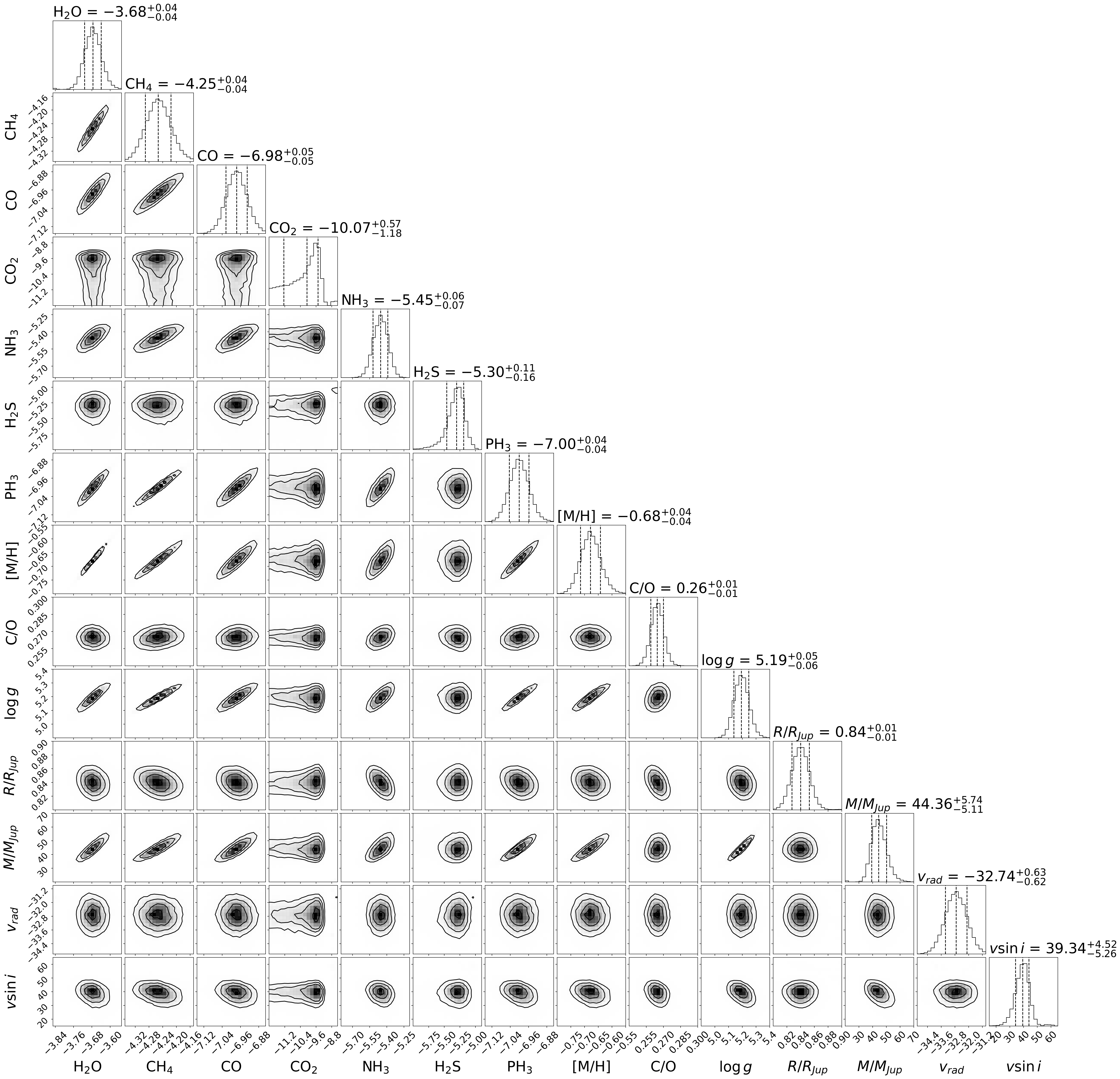}
\caption{{\bf Posterior probability distributions for atmospheric retrieval analysis.}
%the retrieved and inferred parameters for the NIRSpec/G395H spectrum of {\name}.} 
Panels along the diagonal show histograms of the marginalized posterior distributions for the model parameters; while
off-diagonal panels display the marginalized posterior distributions among parameter pairs as contour plots, indicating correlations between parameters.
%Histograms of the marginalized posteriors of parameters are shown in the panels along the diagonal, while contour plots show the correlations between parameters. 
The dashed lines in the histograms indicate the 16\%, 50\%, and 84\% quantiles for the marginalized distributions.
%\textsuperscript{th}, 50\textsuperscript{th}, and 84\textsuperscript{th} percentiles.
%, with the 68\% confidence interval as the width between the 16\textsuperscript{th} and 84\textsuperscript{th} percentiles. 
%Parameter values listed above are estimated as the median~$\pm1\sigma$. 
Gas abundances are displayed on a logarithmic scale ($\log_{10}{f_\mathrm{i}}$).
%as log$_{10}$(X) values, where X is the gas. 
%\Lbol, \teff, radius, mass, and {[M/H]} 
[M/H], C/O, radius, and mass are not directly retrieved parameters, but are inferred using the gas abundances, flux scaling factor $\alpha$, and {\logg}.}
%along with the predicted spectrum. Our [M/H] is relative to Solar.} 
\label{fig:retrieval3}
\end{figure}

\begin{figure}[h]
\centering
A\hspace{2.9in}B \\ 
\includegraphics[width=0.48\textwidth]{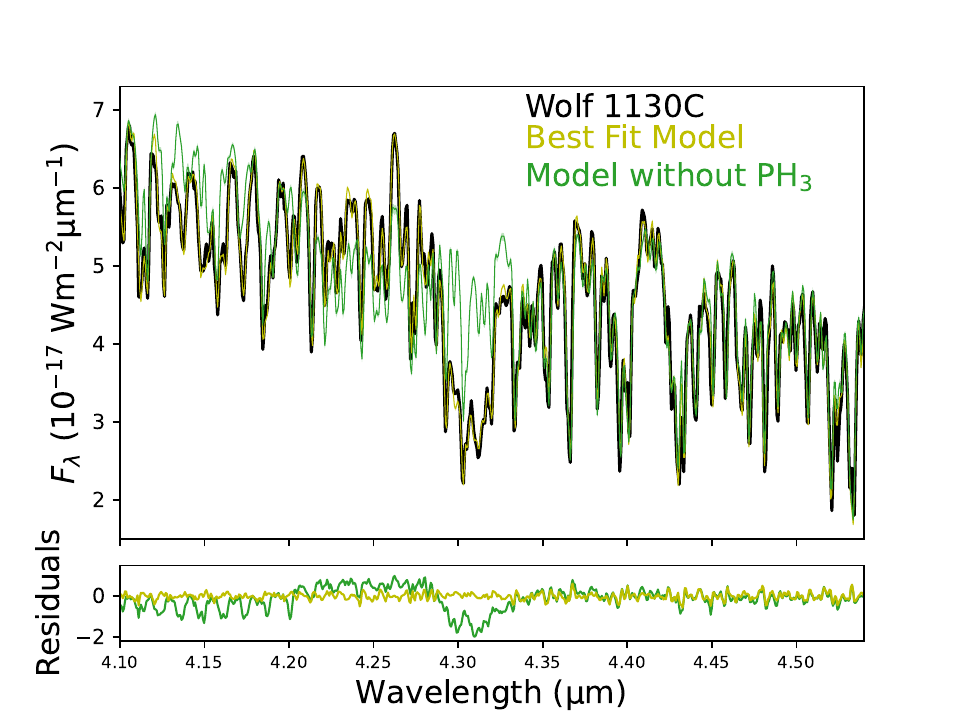} 
\includegraphics[width=0.48\textwidth]{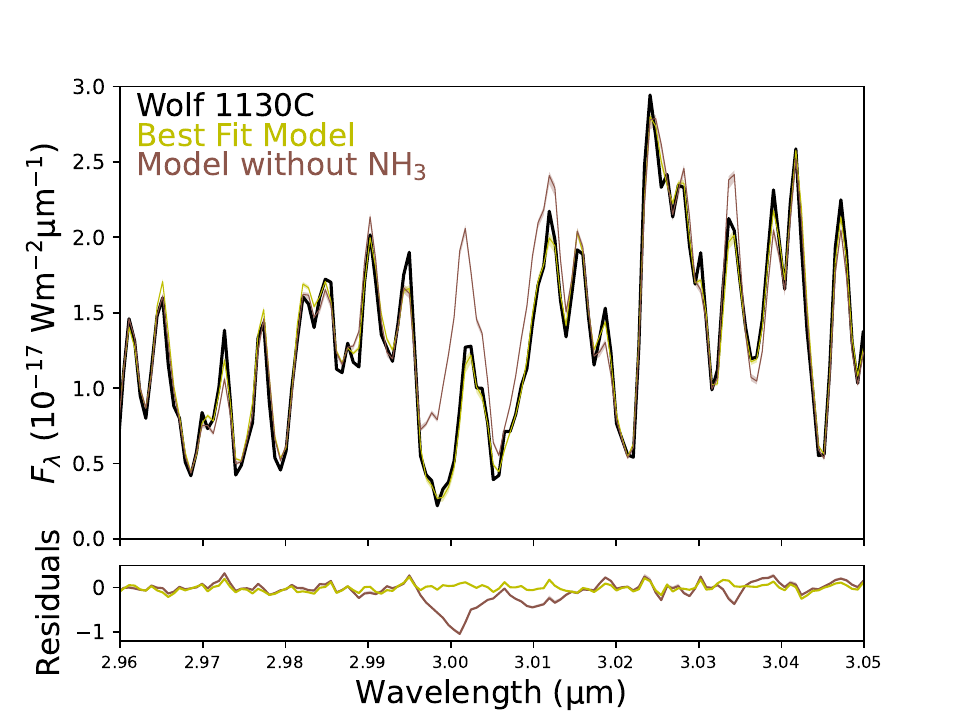} \\
C \\ 
\includegraphics[width=0.48\textwidth]{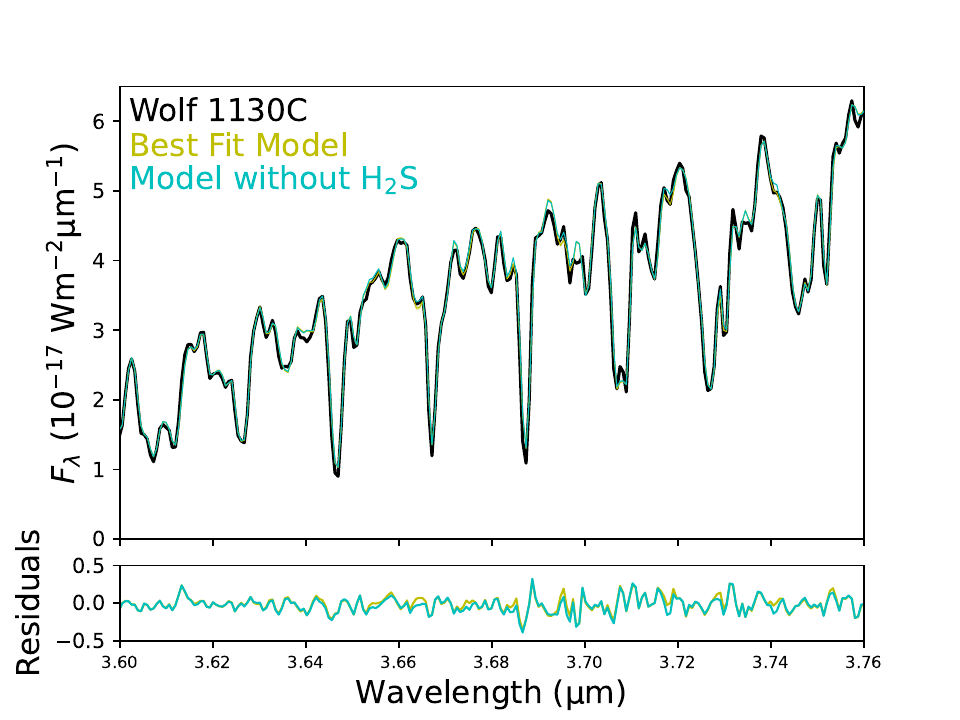} \\
\caption{{{\bf {\phhh}, {\nhhh}, and {\hhs} in the atmosphere of {\name}.}
Each panel shows close-up views of the NIRSpec/G395H spectrum (black line \cite{jwstdata}), the median retrieved model (mustard line),
and the median retrieved model without
{\phhh} (panel A, green line), 
{\nhhh} (panel B, brown line), and 
{\hhs} (panel C cyan line).
At the bottom of each panel is the difference between the 
observed spectrum and the median retrieved model (mustard line) and models with removed molecules (other colored lines).
The Q branches of the $\nu_1$+$\nu_3$ stretch modes of {\phhh} and {\nhhh} are visible at 4.3~{\micron} and 3.0~{\micron}, respectively. 
{\hhs} features in the 3.65 to 3.75~{\micron} region 
%that overlap with {\chhhh} features. This molecule is 
are not discernible in this visual analysis, but its presence is strongly indicated from our retrieval analysis.
}}
\label{fig:nh3-h2s}
\end{figure}

Fig.~\ref{fig:retrieval1} displays the median likelihood spectral model 
drawn from our posterior probability distributions, which reproduces the observed spectrum ($\chi_r^2$ = 6). 
%{Fig.~\ref{fig:retrieval2} displays the retrieved pressure-temperature profile.}
Fig.~\ref{fig:retrieval3} displays the parameter distributions for our retrieved gas abundances, {\logg}, and radial velocity ($v_\mathrm{rad}$).
For our derived bulk metallicity, C/O ratio, radius, and mass are listed in Table~\ref{tab:fit_results} and shown in Figure~\ref{fig:retrieval3}.
The parameters are constrained for all the molecules except 
{\coo}, for which the posterior probability distribution shows a peak at a fractional abundance of $\log{f_\mathrm{CO_2}}$ = $-9.5$ but with a long tail toward the lower limit of our priors; we therefore consider a 3$\sigma$ upper limit of $\log{f_\mathrm{CO_2}}$ = $-9.2$ (0.6 ppb).
We ran separate models for {\name} to test the impact of excluding {\phhh}, {\nhhh}, and {\hhs} absorption,
using 35,000-65,000 iterations with 16 walkers per parameter. Compared to the nominal model that includes all three molecules,
the {\phhh}-free model had a relative $\Delta{BIC}$ = 1,834,
the {\nhhh}-free model had a relative $\Delta{BIC}$ = 212, and
the {\hhs}-free model had a relative $\Delta{BIC}$ = 6,
indicating very strong evidence for {\phhh} and {\nhhh} and strong evidence for {\hhs} based on the criteria of \cite{kass1995}.
Fig.~\ref{fig:nh3-h2s} compares these models to the full model and spectral data in the regions of the molecules' strongest absorption features, which visually confirm the statistical evidence for {\phhh} and {\nhhh}.
%In all three cases we can formally confirm the presence of these molecules . }

%{We also verified the presence of {\phhh}, {\nhhh} and {\hhs} by visually confirming the presence of absorption features in the 3.0~{\micron} and 3.7~{\micron} regions, respectively (Fig.~\ref{fig:nh3-h2s}).}
Assuming the retrieved molecules contain the majority of their respective elements (excluding hydrogen), as motivated by chemical equilibrium models \cite{2021ApJ...920...85M,2006ApJ...648.1181V},
we estimated
elemental abundances relative to the Sun \cite{2003ApJ...591.1220L}.
For carbon, we used the {\chhhh} and {\co} mixing fractions;
for oxygen, we used the {\hho} and {\co} mixing fractions, 
for nitrogen, we used the {\nhhh} mixing fraction, 
for sulfur, we used the {\hhs} mixing fraction, and 
for phosphorus, we used the {\phhh} mixing fraction.
The resulting elemental abundances are listed in Table~\ref{tab:fit_results}.
For nitrogen, our abundance estimate is a lower limit 
because we cannot constrain the disequilibrium abundance of nitrogen locked up in N$_2$ due to vertical mixing \cite{2006ApJ...647..552S}. 
We estimated an overall metal abundance of [M/H] = $-0.68\pm0.04$
by combining all the elemental abundances excluding nitrogen.
We also estimated an alpha element enrichment of [$\alpha$/M] = $+0.30\pm0.04$ as a weighted average of the oxygen and sulfur abundances.
In addition, we estimated C/O = $0.26\pm0.01$ from the relative C and O abundances, significantly lower than the solar ratio of C/O = 0.458 \cite{2003ApJ...591.1220L}.
The overall metallicity and alpha enrichment values of {\name} are consistent with those measured for the atmosphere of the companion star Wolf~1130A \cite{2006PASP..118..218W,2009ApJ...704..975J,2018ApJ...854..145M} and for other metal-poor thick disk stars \cite{2004A&A...414..931A}.

From the retrieved {\logg} and radius $R$, the latter derived from the scaling factor $\alpha = (R/d)^2$ (Eqn.~\ref{eqn:alpha}), 
%$(R/d)^2$, allows us to directly 
we infer the mass of {\name} to be {M = $gR^2/G$ =} 44$^{+6}_{-5}$~{\mjup}, where $G$ is the universal gravitational constant. This mass is consistent with an old, evolved brown dwarf, as also indicated by 
%($\tau \gtrsim 10$~Gyr) 
the measured luminosity and {inferred} {\teff} from our SED analysis.
%\cite{2003A&A...402..701B,2001RvMP...73..719B,2020A&A...637A..38P,2021ApJ...920...85M,2024ApJ...971...65G}.
Our retrieved $v_\mathrm{rad}$ is {also} consistent with the center of mass motion of the Wolf~1130AB binary system,  which has previously been measured as $-$33.2$\pm$0.2~km~s$^{-1}$ \cite{2018ApJ...854..145M}.
%We defer discussion of the retrieved {\vsini} to a later study as {its value is entangled with} the currently unconstrained instrumental broadening profile.

\begin{figure}[h]
\centering
A \\ \vspace{0.2cm} 
\begin{minipage}{\textwidth}
\hspace{2.5cm}
\includegraphics[width=0.65\textwidth]{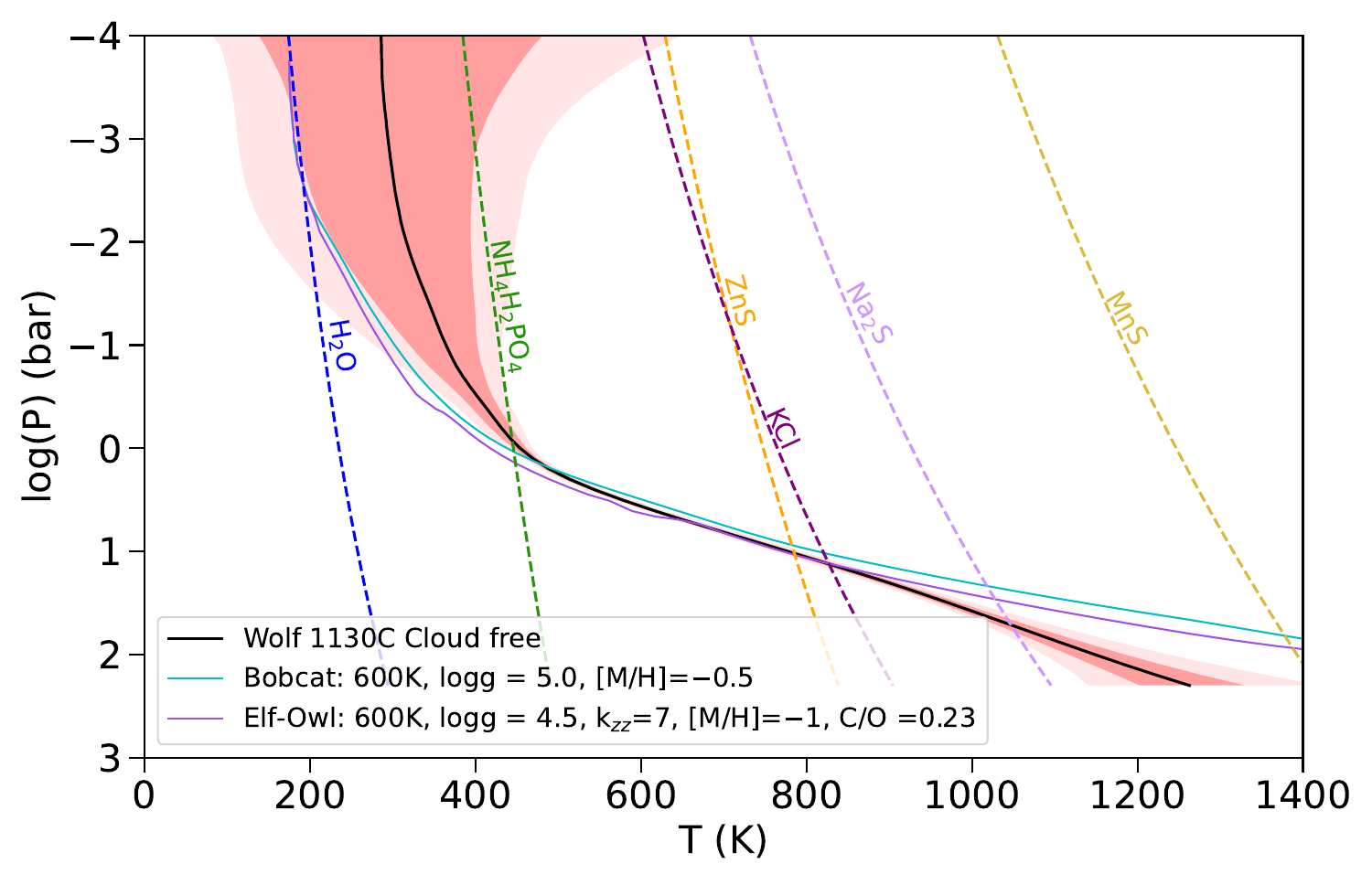} \hspace{1cm} \end{minipage}
\\ 
B \\ \vspace{0.2cm}  \includegraphics[width=0.7\textwidth]{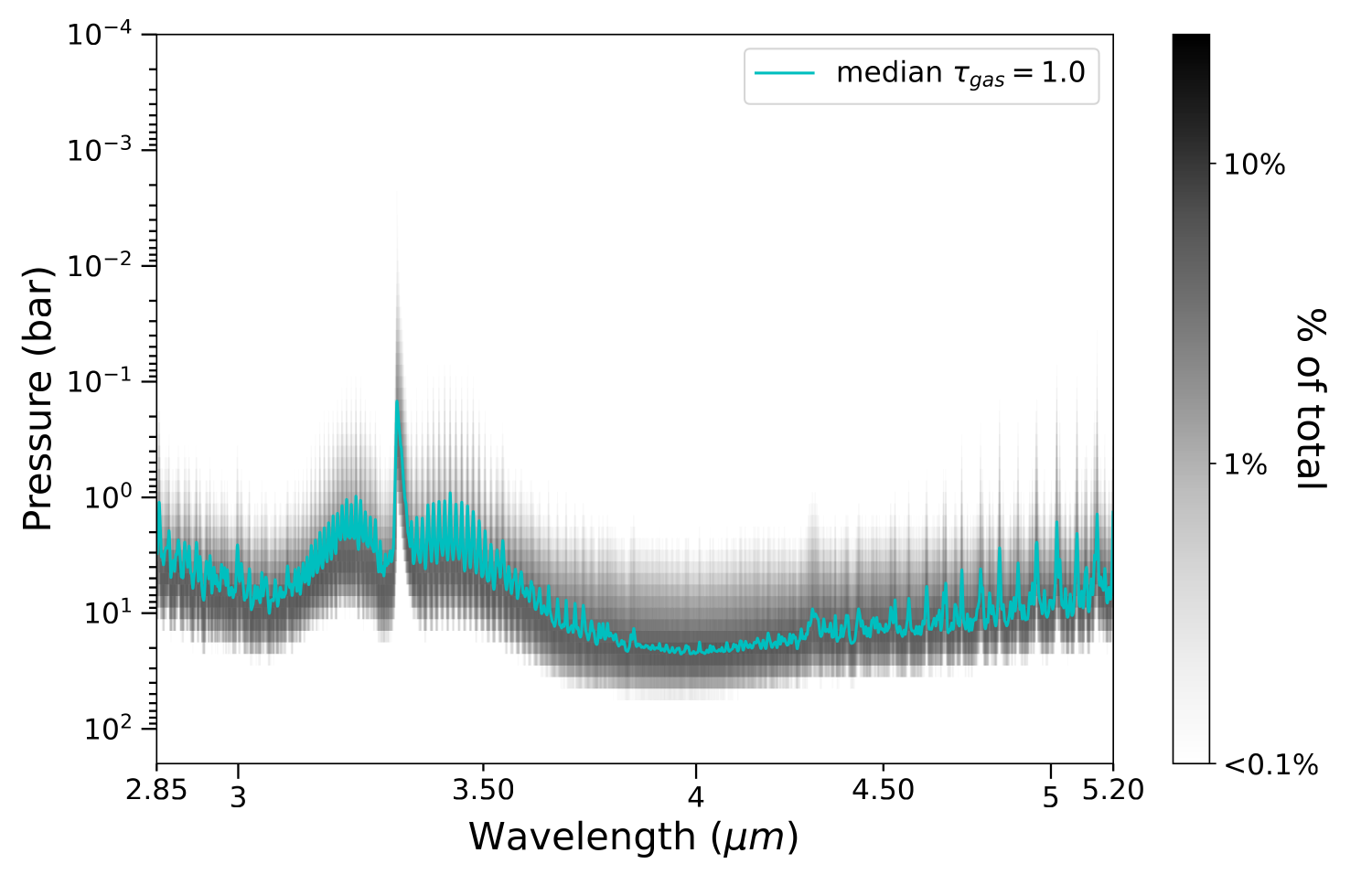} \\
\caption{{\bf The retrieved thermal profile and contribution function for {\name}.} 
(A) The median thermal profile from our retrieval analysis (black line), with $\pm$1 and $\pm$2 sigma confidence intervals (dark and light salmon shading), respectively). 
Overlain are predicted model pressure/temperature profiles from 
%Sonora Bobcat (blue line \cite{2021ApJ...920...85M}) and Elf-Owl models (purple line \cite{2024ApJ...963...73M}) 
Sonora Bobcat (blue line) and Elf-Owl (purple line) models
for the parameters listed in the legend. Dashed lines are the
condensation curves at [M/H] = $-$0.68 for {\hho}, {\adp}, ZnS, KCl, Na$_2$S, and Mn,S as labeled.
%(labeled and colored dashed lines).
% (Top panel) {The NIRSpec/G395H spectrum of {\name} (black line \cite{jwstdata}) compared to the median likelihood atmosphere model from our retrieval analysis (mustard line),
% which is nearly indistinguishable with a reduced $\chi^2_r$ = 6.
% Single pixel deviations at 5.06~{\micron} and 5.14~{\micron} are due to noise.} 
(B) The contribution function {to the retrieved spectrum of {\name} at each wavelength (black shaded region). The} median pressure level at a gas optical depth of $\tau_{gas}=1$ is overplotted (aqua line).
The spike and broad feature near 3.3~{\micron} arises from {\chhhh} absorption.
{The bulk of emission emerges from the 1--20~bar pressure region.}}
\label{fig:retrieval2}
\end{figure}

Fig.~\ref{fig:retrieval2}A displays the retrieved thermal profile of {\name}. Above a pressure of $\sim$10~bar, this profile is consistent with the thermal profiles of metal-poor, {\teff} = 600~K atmospheres predicted by the 
Sonora Bobcat \cite{2021ApJ...920...85M} and Elf-Owl \cite{2024ApJ...963...73M} models. However, the retrieved profile deviates in the deep photosphere, becoming colder at a given pressure as compared to the adiabatic profiles assumed for the grid models. 
%Similarly, we see a marginally significant deviation in the upper photosphere, with the retrieved profile shifting to warmer temperatures relative to the grid models.
This shallower temperature gradient in the retrieved thermal profile could be due to condensation of KCl and ZnS in the deep photosphere.
%and {\adp} at the upper photosphere \cite{2012ApJ...756..172M}. 
There is no evidence of continuum opacity or element depletion associated with these condensates, so if they are present it would be in the form of 
%their possible influence on the thermal profile may indicate the presence of 
optically thin cloud layers. 
This deviation occurs at depths that are near the limits of where are retrieval modeling is sensitive to the thermal profile. The sensitivity range is indicated by the contribution function 
(Fig.~\ref{fig:retrieval2}B) which maps
%The retrieved thermal profile does not intersect with the {\hho} condensation line, consistent with the relatively high {\hho} gas abundance retrieved in our analysis. 
%Figure~\ref{fig:retrieval3} also shows the contribution function in this spectral region, indicating 
the pressure layer {from which the observed spectrum emerges}
as a function of wavelength. 
{The deepest layers at $\sim$30~bar are probed around the 4~{\micron} peak, the region encompassing {\phhh} absorption,} which is considerably deeper in {\name} than in solar-metallicity brown dwarfs \cite{2024Natur.628..511F}, and consistent with a more transparent atmosphere with less molecular gas opacity.
{The majority of the observed spectrum emerges from the 1--20~bar pressure region, where our retrieved profile is in good agreement with that predicted by the metal-poor Elf-Owl model grid.}

\subsubsection*{Chemical Abundance Analysis\label{sec:chemeq}}

To explore the role of {mixing, elemental abundances, and thermodynamic assumptions on the atmospheric phosphorus and carbon chemistry} of {\name}, we calculated a suite of chemical equilibrium abundances for {\phhh} and {\coo}.
%the {primary} molecular species observed in the 4--5~{\micron} region of cool brown dwarf spectra. These calculations 
We followed the approach of previous {studies} \cite{2021ApJ...920...85M,2006ApJ...648.1181V}, {with} equilibrium abundances computed over a wide range of pressures and temperatures {at} select {metal abundances ([M/H])} and C/O {abundance} ratios.  Fig.~\ref{fig:chemistry} displays the equilibrium abundances
{for {\phhh} for both solar
%{associated with phosphorus and carbon chemistry, specifically} PH$_3$ and {\coo}, 
%as a function of temperature and pressure 
%for solar {
and one-tenth solar metal abundances and assuming}
%of [M/H] = 0.0 and $-$1.0 dex, the latter assuming atmospheric 
C/O = 0.3. We also show the retrieved thermal profile for {\name} {in} this pressure-temperature space. 
These calculations confirm prior studies that predict {\phhh} contains the
bulk of the phosphorus reservoir in the photospheres of cool brown dwarfs in equilibrium conditions \cite{2006ApJ...648.1181V}.
%(a mixing ratio of $\sim$3~ppm assuming an elementa abundance of [P/H] = $-$6.5 \cite{})  of just over 2~ppm, 
In the metal-poor case of {\name}, our model predicts a mixing fraction of 50~ppm along the thermal profile at pressures of 5--20~bar.
%
%As for other cool brown dwarf and gas giant atmospheres, 

In these models, PH$_3$ {remains the} dominant {phosphorus}-bearing gas until {it is removed} by oxidation and/or condensation at low temperatures. The chemical products of those processes remain
%The identity of the low-temperature oxide or condensate remains 
poorly constrained due to uncertainties in the enthalpies of formation of oxides such as P$_4$O$_6$
{\cite{1995Icar..113..460B,2016Icar..276...21W,2023ESC.....7.1219B,2024ApJ...976..231L}.}
Different oxide phases have been considered {in the literature,} including 
P$_4$O$_6$, 
P$_4$O$_{10}$, {and} 
H$_3$PO$_4$ {gases}, and {the condensate} 
NH$_4$H$_2$PO$_4$ \cite{1994Icar..110..117F,1995Icar..113..460B,2006ApJ...648.1181V,2016Icar..276...21W}. 
{To explore uncertainties in phosphorus chemistry, we calculated equilibrium abundances for two sets of enthalpy data for P$_4$O$_6$: 
a nominal chemistry \cite{1989hpc..book.....G,Mcbride2002NASAGC} in which {\phhh} is replaced by H$_3$PO$_4$ vapor and then removed by NH$_4$H$_2$PO$_4$ condensation at {\teff} $\approx$ 500~K \cite{2016Icar..276...21W,2020JGRE..12506526V,2023ESC.....7.1219B};
and a modified chemistry based on P$_4$O$_6$ thermodynamic data from the Joint Army-Navy-Air Force (JANAF) thermodynamic tables \cite{chase1998} in which {\phhh} is replaced first by P$_4$O$_6$ followed by condensation into NH$_4$H$_2$PO$_4$ \cite{2006ApJ...648.1181V}.
These two scenarios are illustrated in Fig.~\ref{fig:chemistry}, and indicate
that -- in the case of chemical equilibrium -- the nominal chemistry retains {\phhh} in the photosphere of {\name}, whereas the modified chemistry results in a depletion of {\phhh} to ppb levels, which is consistent with measurements and upper limits for other brown dwarfs and gas giant exoplanets \cite{2024ApJ...973...60B,2024ApJ...971..121K,2024arXiv240205345H,2024ApJ...977L..49R}.
Hence, while a change in P$_4$O$_6$ thermodynamic quantities could explain the depletion of {\phhh} in other low temperature atmospheres, it does not explain the abundance of {\phhh} in the atmospheres of {\name}, Jupiter, or Saturn.
%argues against such fundamental shift in phosphorus thermodynamics. 
In either case, vertical mixing of {\phhh} from deep in the atmosphere (where it is most abundant)
rules out variations in equilibrium chemistry as an explanation for this molecule's depletion.
%at rates faster than chemical conversion timescales, and should roughly approximate the atmospheric phosphorus inventory \cite{2020JGRE..12506526V}.
%well into the upper atmosphere. 
% Indeed, it is the expectation of a mixing enhancement 
% is why the lack of observed {\phhh} features in cool brown dwarf and exoplanet atmospheres to date remains a puzzle. 
The abundant {\phhh} in the atmosphere of {\name} is compatible with predictions of nominal chemistry models if they also include vertical mixing.
%current thermochemistry and vertical mixing chemistry models. 
It remains unclear why {\phhh} has such a low abundance in other brown dwarf and giant exoplanet atmospheres.
%, making the lack of this feature in other brown dwarf and giant exoplanet spectra all the more mysterious.
}
% enrichment previously measured in the atmospheres of Jupiter and Saturn \cite{1985ApJ...299.1067F,2005ApJ...623.1221V}.}

% Here we adopt formation enthalpy data for P$_4$O$_6$ from the compilation of \cite{1989hpc..book.....G}. In this scenario, PH$_3$ is the dominant gas until its gradual replacement by H$_3$PO$_4$ vapor and then removal by NH$_4$H$_2$PO$_4$ condensation at {\teff} $\approx$ 500~K, indicated on the left-hand side of Figure~\ref{fig:chemistry}. Concurrently, PH$_3$ is expected to undergo quenching deep in the photosphere where it is relatively abundant, so PH$_3$ abundances through the photosphere roughly approximate the atmospheric phosphorus inventory.
% %well into the upper atmosphere. 
% The expectation of a mixing enhancement 
% is why the lack of observed {\phhh} features in cool brown dwarf and exoplanet atmospheres to date has been an ongoing puzzle. 
% The detection of {\phhh} in the spectrum of {\name} is more in line with expectations based on the enrichment previously measured in the atmospheres of Jupiter and Saturn \cite{1985ApJ...299.1067F,2005ApJ...623.1221V}.

\begin{figure}[h]
\centering
% A\hspace{3.1in}B \\ 
% \vspace{0.2cm}
%\includegraphics[width=0.95\textwidth]{CO2-plot-revision.pdf} \\
 %A \\ ~\\ 
 \includegraphics[width=0.5\textwidth]{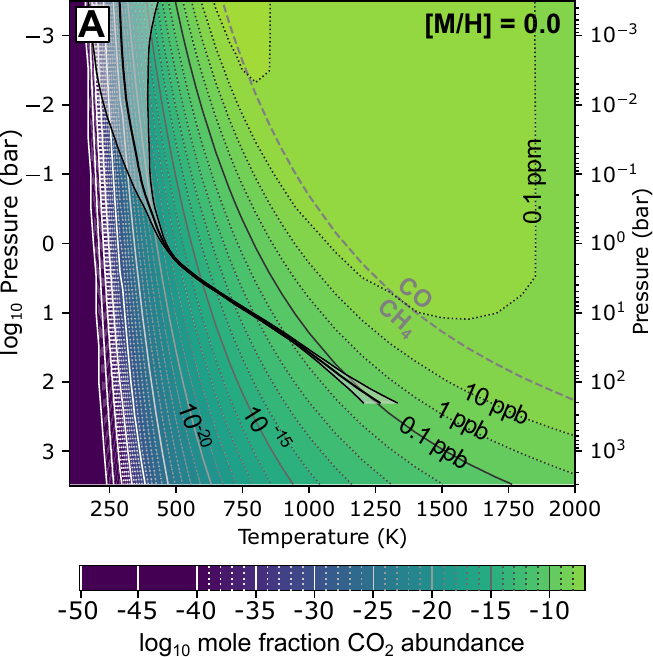} \\
 %B \\ ~\\ 
 \vspace{1cm}
 \includegraphics[width=0.5\textwidth]{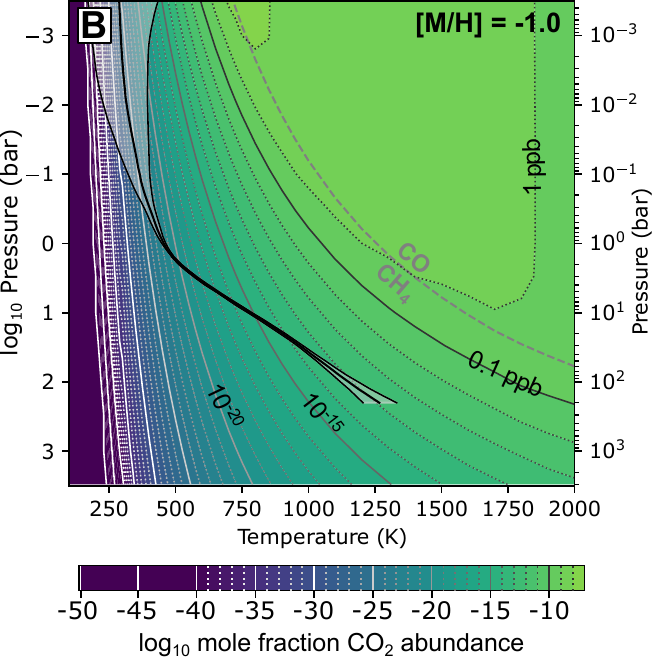}
\caption{{{\bf Chemical equilibrium abundances of {\coo} in the atmosphere of {\name}.}
Same as Fig.~\ref{fig:chemistry} but for {\coo}.
(A) Results for gas mixtures with 
C/O = 0.3 and solar metal abundances ([M/H] = 0).
(B) Results for and one-tenth solar metal abundances ([M/H] = $-$1).
Abundance values are indicated by the color bar, with isoabundance lines indicated in 1~dex (dotted) and 5~dex (solid) increments (black on green, white on purple) and labeled at select intervals.
The grey dashed line indicates the equal abundance curve for the primary carbon-bearing species CO and {\chhhh}
}}
\label{fig:chemistry2}
\end{figure}

We also calculated the equilibrium abundance of {\coo} using the same chemical models, with the results shown in Fig.~\ref{fig:chemistry2}. 
This model predicts that {\chhhh} is the dominant {carbon}-bearing gas throughout {the photosphere of {\name},
with {\coo} abundances decreasing toward lower temperatures as indicated in the Figure.
%decrease from 0.1~ppb to trace levels (10$^{-20}$) into the lower-temperature regions of the photosphere. 
%the thermal profile {of }over the temperature range shown, with {\coo} abundances decreasing 
relative to the {\chhhh}-{\co} equal-abundance boundary. 
The model {\coo} abundance is highly sensitive to the C/O ratio and the overall metallicity of the atmosphere \cite{2011ApJ...734...73T,2024ApJ...963...73M}. {While a} low value of C/O (as inferred for {\name}) favors {\coo} formation, the {low overall} metallicity strongly reduces the {\coo} abundance. 
%Given the relative sensitivity to metallicity of {\phhh} and {\coo} \cite{2011ApJ...734...73T,2024ApJ...962..177B}, 
%A shift from [M/H] = 0.0 to $-$1.0 {reduces {\coo} by a factor of $\sim$1,000, and o
Our equilibrium models predict undetectable mixing ratios of 10$^{-14}$ to 10$^{-20}$ across the photosphere of {\name}.}
However, {\coo} {is also vertically mixed,} yielding abundances
much higher than the predictions of equilibrium models
in low-temperature atmospheres \cite{2011ApJ...734...73T,2024ApJ...973...60B}.
%(Figure~\ref{fig:chemistry}). 
Our analysis suggests that for {\name} such mixing is insufficient to raise the photospheric {\coo} abundance enough to {produce the typically prominent $\nu_2$ feature at 4.3~{\micron}.
Metallicity effects on {\phhh} and {\coo} abundances imply a $\sim$100$\times$ increase in the {\phhh}/{\coo} abundance ratio relative to solar elemental abundances regardless of vertical mixing \cite{2011ApJ...734...73T,2024ApJ...962..177B}.}
%at temperatures and pressures where {\chhhh} is the dominant C-bearing gas and PH$_3$ is the dominant P-bearing gas. 

% To refer to this section from the main text, use the numbered note in the reference list \cite{methods}.
% Refer to figures and tables in the same way as in the main text but now all capitalized e.g.
% Fig.~\ref{fig:example}, Table~\ref{tab:example},
% Fig.~\ref{fig:sup_example} and Table~\ref{tab:sup_example}.
% Cite references in the usual way \cite{example2},
% including any that are only cited in the supplement \cite{sm_example,conference_example}.

%%%%%%%%%%%%%%%% SUPPLEMENTARY TEXT %%%%%%%%%%%%%%%

\subsection*{Supplementary Text}
% The Supplementary Text section can only be used to directly support statements made in the main text
% e.g. to present more detailed justifications of assumptions, investigate alternative scenarios,
% provide extended acknowledgements etc.
% Material in this section cannot claim results or conclusions that weren't mentioned in the main text.
% To refer to this section from the main text, just write (Supplementary Text).

% \subsubsection*{Example supplement heading}

% The two main sections of the supplement can be split up using headings.

% If your supplement is very short you might need to uncomment the following line to avoid
% layout problems with the figures and tables.

\subsubsection*{Phosphorus in the Spectrum of Wolf~1130A}

{As discussed in the main text,} the phosphorus in the atmosphere of {\name} could originate in mass transferred from the Wolf~1130B progenitor 
\cite{2005NuPhA.758..259G,2011MNRAS.415.3865V,2020NatCo..11.3759M,2023Natur.623..292K} or ongoing {accretion}-driven nucleosynthesis
\cite{1998ApJ...494..680J,2014MNRAS.442.2058D,2024ApJ...967L...1B}.
Any mass transfer would leave a {stronger} imprint on the spectrum of the closer Wolf~1130A primary star as compared to {\name}. While the atmosphere of Wolf~1130A is too warm for {\phhh} formation ({\teff} = 3530$\pm$60~K \cite{2018ApJ...854..145M}), atomic phosphorus is found in the near-infrared spectra of metal-poor phosphorus-rich M giants with comparable temperatures \cite{2022A&A...668A..88N,2023A&A...673A.123B}. We examined previously-acquired high resolution ({\ldl} = 45,000) near-infrared spectral data for Wolf~1130A 
%obtained with the Immersion Grating Infrared Spectrometer (IGRINS \cite{2014SPIE.9147E..1DP}) reported in 
\cite{2018ApJ...854..145M}.
The spectrum has a signal-to-noise of 500--1,000 around the 1.57115~{\micron} and 1.63829~{\micron} P~\textsc{i} lines  \cite{1980PhyS...22..288S,2018ApJ...865...44A,2022A&A...668A..88N}. 
%Figure~\ref{fig:p1-1130} displays this spectrum around these P~I transitions.
We find no evidence of P~\textsc{i} absorption and set 5$\sigma$ upper limits on the equivalent widths of 7 m$\overset{\circ}{\mathrm{A}}$ and 2 m$\overset{\circ}{\mathrm{A}}$, respectively. We conclude there is no evidence for phosphorus-rich mass transfer from Wolf~1130B to Wolf~1130A, so no reason to expect it occurred for Wolf 1130C.
% , {and hence} no {compelling} evidence
% %of phosphorus enrichment in this star, arguing 
% that the modest enrichment in {\name} {arises} from pollution from Wolf~1130B or its progenitor. 

\clearpage

%%%%%%%%%%%%%%%% SUPPLEMENTARY FIGURES %%%%%%%%%%%%%%%

% \begin{figure}[h]
% \centering
% \includegraphics[width=0.75\textwidth]{mcmcfit_Wolf1130C_JWST-NIRSPEC-PRISM_elfowl24_compare2.pdf} \\ A \\
% \includegraphics[width=0.75\textwidth]{mcmcfit_Wolf1130C_JWST-NIRSPEC-PRISM_elfowl24-ph3_compare2.pdf} \\ B \\
% %\includegraphics[width=0.6\textwidth]{mcmcfit_Wolf1130C_JWST-NIRSPEC-PRISM_elfwol24-ph3_corner2.pdf} 
% \caption{{\bf
% Sonora Elf-Owl model fits to the NIRSpec/PRISM spectrum of {\name}}.
% {Panel A displays fits using the published models with artificially suppressed {\phhh} abundances (EOwl \cite{2024ApJ...963...73M}); 
% panel B displays fits using models with {\phhh} abundances
% set by vertical mixing (EOwl+ \cite{2024ApJ...973...60B}). 
% Both figures compare the} absolute flux-calibrated spectrum (black line \cite{jwstdata}) to the best-fit model (magenta line) and {20} draws from the MCMC posterior chains (semi-transparent {magenta} lines). {Sub-panels show the difference between observed and best-fit model spectra relative to the {$\pm$5$\sigma$} uncertainty (grey bands).}
% %(Bottom panel) Marginalized posterior distributions of model parameters {\teff}, {\logg}, [M/H], C/O, $\log\kappa_{zz}$, and radius along the diagonal histograms, and parameter correlations among the inner counter plots. The values delineated by the magenta lines correspond to the overall best-fit model parameters, while the dashed lines in the posterior distributions indicate quantiles of 16\%, 50\% and 84\%.
% }
% \label{fig:gridfit}
%\end{figure}

% \begin{figure}[h]
% \centering
% \includegraphics[width=0.48\textwidth]{wolf1130c_G395H_nh3.pdf}
% \includegraphics[width=0.48\textwidth]{wolf1130c_G395H_h2s.pdf} \\
% \caption{{{\bf Confirmation of the presence of {\nhhh} and {\hhs} in the atmosphere of {\name}.}
% Both panels display close-up views of the NIRSpec/G395H spectrum (black line \cite{jwstdata}) compared to opacity spectra for
% {\hho} (blue \cite{2018MNRAS.480.2597P})
% {\chhhh} (orange \cite{10.1093/mnras/stae148})
% {\nhhh} (brown \cite{10.1111/j.1365-2966.2011.18261.x}), and
% {\hhs} (cyan, scaled by a factor of 2 \cite{10.1093/mnras/stw1133}).
% Absorption spectra are computed based on the photosphere temperature and 
% relative abundances for inferred from our retrieval analysis, and offset by a constant. 
% The Q branch of the {\nhhh} $\nu_1$+$\nu_3$ stretch mode is labeled in the left panel (horizontal bar), while weak features from
% {\hhs} that thread through {\chhhh} opacity
% in the 3.65--3.75~{\micron} region are labeled in the right panel (vertical dashed lines).
% }}
% \label{fig:nh3-h2s}
% \end{figure}

\clearpage

%%%%%%%%%%%%%%%% SUPPLEMENTARY TABLES %%%%%%%%%%%%%%%

\begin{table}[h]
\caption{{\bf Parameters and priors for the retrieval analysis. }
The first column {provides} a descriptor for each retrieval parameter,
the second column shows the corresponding symbol used in the text and Fig.~\ref{fig:retrieval3},
and the third column lists the assumed distribution {type} and range of the priors. 
The gas number fractions were fit for molecules
{\hho}, {\chhhh}, {\co}, {\coo}, {\nhhh}, {\hhs} and {\phhh}. \\
}
\label{tab:priors}
\begin{tabular}{lcl}
\hline
{\bf Parameter} & {\bf Variable} & {\bf Prior} \\
\hline
Gas number fraction & $f_{i}$ & log-uniform, $\log f_{i} \geq -12.0$, $\sum_{i}{f_{i}} \leq 1.0$ \\
Thermal profile & $T(P)$ & uniform, constrained by $0~\mathrm{K} < T < 5,000~\mathrm{K}$ \\
Profile smoothing & $\gamma$ & uniform, $0 < \gamma < 10^5$ \\
Tolerance factor & $b$ & uniform, $\log (0.01 \times \mathrm{min}(\sigma_{i}^{2})) \leq b \leq \log(100 \times \mathrm{max}(\sigma_{i}^{2}))$ \\
Surface gravity & {\logg} &  uniform, constrained by 1~{\mjup}  $\leq gR^{2} / G \leq$ 80~{\mjup}  \\ 
Scale factor &  $\alpha = \left(R/d\right)^2$ & uniform, assumes $d$ = 10~pc and 0.5~{\rjup} $\leq$ $R$ $\leq$ 2.0~{\rjup} \\
Rotational velocity & {\vsini} & uniform, 0~km~s$^{-1}$ $<$ {\vsini} $<$ 150~km~s$^{-1}$ \\
Radial velocity & $v_{rad}$ & uniform, $-$250~km~s$^{-1}$ $<$ $v_\mathrm{rad}$ $<$ 250~km~s$^{-1}$ \\
\hline
\end{tabular}
\end{table}

\newpage

\begin{table}[h]
\centering
\caption{{\bf Derived atmosphere parameters {and abundances} for {\name}.}
{The top section} lists physical parameters
{inferred from}
% logarithmic solar-scaled luminosity ($\log(\mathrm{L_{bol}/L_\odot})$),
% effective temperature ({\teff} in units of $\degree$K),
% and radius (in units of Jupiter radii, {\rjup})
% based on 
the SED fitting analysis.
%(Sec.~$\ref{sec:sedfit}$)
The second section lists physical parameters 
inferred from the Sonora Elf Owl models with {\phhh}
abundances set by vertical mixing (EOwl+ \cite{2024ApJ...973...60B}) fitted 
to the NIRSpec/Prism spectrum; see Fig.~\ref{fig:gridfit}.
% effective temperature,
% log surface gravity ({\logg} in units of cm/s$^2$),
% solar-scaled metallicity ([M/H] in units of dex relative to the Sun),
% C/O abundance ratio (linear scaling, where C/O$_\odot$ = 0.458 \cite{2003ApJ...591.1220L}),
% log vertical mixing diffusion coefficient ($\log\kappa_{zz}$ in units of cm$^2$/s),
% and radius.}
{The third section lists} 
physical parameters inferred from our retrieval analysis.
% (Sec.~\ref{sec:retrieval}),
% surface gravity,
% radial velocity ($v_r$ in units of km/s)
% radius (in units of {\rjup}), and
% mass (in units of Jupiter masses {\mjup}).
{The bottom section} lists the molecular abundances 
{of {\hho}, {\chhhh}, {\co}, {\coo}, {\nhhh}, {\hhs} and {\phhh}}
inferred from the retrieval analysis, 
reported as volume mixing ratios assumed to be constant throughout the modeled photosphere; 
and the derived elemental abundances relative to solar,
{assuming} elements are fully contained in the retrieved molecules.
%(note that an unknown fraction of N may be contained in undetected N$_2$)}. 
%For nitrogen, we assume this value is a lower limit given the unknown abundances of $N_2$. 
{We also list the bulk metallicity} [M/H] based on all of the metal abundances excluding nitrogen, the alpha element abundance ratio [$\alpha$/M] based on the uncertainty-weighted average of [O/M] and [S/M], {and the C/O ratio based on the carbon and oxygen abundances.}
{All} uncertainties reflect 68\% confidence intervals.
Upper limit for {\coo} reflects 3$\sigma$ limit.
Lower limit for [N/M] reflects the unknown abundance of N contained in N$_2$. 
Where multiple values are available from different analyses, those marked with a $^*$ are our assumed to be fiducial values.\\
}
\label{tab:fit_results}
\begin{tabular}{lcc|lcc}
\hline%
\multicolumn{6}{l}{\bf SED Physical Parameters} \\
%\hline
\bf{Quantity}& 
\bf{Value}& 
\bf{Unit} &
\bf{Quantity}& 
\bf{Value}& 
\bf{Unit}\\
\hline
$\log(\mathrm{L_{bol}/L_\odot})$ & $-$6.047$\pm$0.003 & dex & Radius$^*$ & 0.80$\pm$0.02 & {\rjup} \\
{\teff}$^*$ & 621$\pm$9 & $\degree$K \\
\hline
% \end{tabular}
% \begin{tabular}{lcc}
% \hline
\multicolumn{6}{l}{\bf Grid Model Fit Physical Parameters}\\
%\hline
\bf{Quantity}& 
\bf{Value}& 
\bf{Unit} &
\bf{Quantity}& 
\bf{Value}& 
\bf{Unit}\\
\hline
{\teff} & 593$^{+21}_{-18}$~K & $\degree$K & C/O & 0.32$^{+0.09}_{-0.04}$ & \\
{\logg} & 4.9$\pm$0.3  & cm~s$^{-2}$ & $\log\kappa_{zz}$ & 6.5$^{+1.6}_{-1.9}$ & cm$^2$~s$^{-1}$ \\
{[M/H]} &  $-$0.93$^{+0.11}_{-0.07}$ & dex & Radius & 0.74$^{+0.04}_{-0.02}$ & {\rjup} \\
%EOwl $\chi^2_r$ & 12.0 &  & EOwl+ $\chi^2_r$ & 9.4 &  \\
\hline
\multicolumn{6}{l}{\bf Retrieved Physical Parameters}\\
% \multicolumn{3}{l}{\bf Retrieved Physical Parameters} & 
% \multicolumn{3}{l}{\bf Derived Physical Parameters}\\
%\hline
\bf{Quantity}& 
\bf{Value}& 
\bf{Unit}& 
\bf{Quantity}& 
\bf{Value}& 
\bf{Unit} \\
\hline
$\log g$$^*$ & 5.19$^{+0.05}_{-0.06}$ & dex & Radius & 0.84$\pm$0.01 & {\rjup}\\
$v_{rad}$ & $-$32.7$\pm$0.7 & km/s & Mass & 44$^{+6}_{-5}$ & {\mjup} \\
%\hline%
\hline
\multicolumn{6}{l}{\bf Retrieved Abundances}\\
% \multicolumn{3}{l|}{\bf Retrieved Molecular Abundances} & 
% \multicolumn{3}{l}{\bf Derived Elemental Abundances}\\
\bf{Species}& 
\bf{Log Abundance}& 
\bf{Unit} &
\bf{Quantity}& 
\bf{Value}& 
\bf{Unit}\\
\hline
{\hho} & $-3.68\pm0.04$ & dex & {[C/M]} & $+0.04\pm{0.06}$ & dex\\
{\chhhh} & $-4.25\pm0.04$ & dex & {[O/M]} & $+0.32\pm{0.06}$ & dex \\
{\co} & $-6.98\pm0.05$ & dex & {[N/M]} & $\gtrsim$$-0.60\pm{0.13}$ & dex \\
{\coo} & $\lesssim-9.2$ & dex & {[S/M]} & +0.19$^{+0.13}_{-0.16}$ & dex\\
{\nhhh} & $-$5.45$^{+0.06}_{-0.07}$ & dex & {[P/M]} & $+0.22\pm{0.06}$ & dex\\
{\hhs} & $-$5.30$^{+0.11}_{-0.16}$ & dex & {[M/H]}$^*$ & $-0.68\pm{0.04}$ & dex\\
{\phhh} & $-7.00\pm0.04$ & dex & {[$\alpha$/M]} & $+0.30\pm0.04$ & dex \\
 & & & C/O$^*$ & $0.26\pm{0.01}$\\
\hline
% \end{tabular}
% \begin{tabular}{lcc}
% \hline
\end{tabular}
\end{table}

%%%%%%%%%%% CAPTIONS FOR OTHER SUPPLEMENTARY FILES %%%%%%%%%%

%\clearpage % Clear all remaining figures and tables then start a new page

% \paragraph{Caption for Movie S1.}
% \textbf{All captions must start with a short bold sentence, acting as a title.}
% Then explain what is shown in the supplementary video file.
% Give as much detail as you would for a figure e.g. explain axes, color maps etc.
% If the video is an animated equivalent of one of the static figures, state e.g.
% `Animated version of Figure~\ref{fig:example}.'

% \paragraph{Caption for Data S1.}
% \textbf{Spectral Data for Wolf~1130C and Ross~458C.}
% % Then explain what is included in the supplementary data file.
% Give as much detail as you would for a table e.g. explain the meaning of every column,
% units used, any special notation etc.

%%%%%%%%%%%%%%%% SUPPLEMENTARY REFERENCES %%%%%%%%%%%%%%%

% Do NOT include a reference list in the supplement.
% All references must be in a single list at the end of the main text.
% The copyeditors will ensure that the correct reference list appears with each version of the paper
% (print, HTML, PDF, mobile app, metadata for bibliographic databases etc.)

\end{document}